\documentclass[9pt,oneside,twocolumn]{IEEEtran}
\usepackage{cite}

\usepackage{cprotect}

\ifCLASSINFOpdf
  \usepackage[pdftex]{graphicx}
  \graphicspath{{../pdf/}{../jpeg/}}
  \DeclareGraphicsExtensions{.pdf,.jpeg,.png}
\else
  \usepackage[dvips]{graphicx}
 \graphicspath{{../eps/}}
  \DeclareGraphicsExtensions{.eps}
\fi

\usepackage[cmex10]{amsmath}
\usepackage{amssymb}
\usepackage{algorithmic}

\usepackage{array}

\ifCLASSOPTIONcompsoc
\usepackage[caption=false,font=normalsize,labelfont=sf,textfont=sf]{subfig}
\else
\usepackage[caption=false,font=footnotesize]{subfig}
\fi
\usepackage{stfloats}
\usepackage{url}

\usepackage{mathtools}

\usepackage{setspace}
\usepackage{lineno}
\usepackage{xcolor}

\begin{document}
%
\title{A Novel No-reference Video Quality Metric for \\Evaluating Temporal Jerkiness due to Frame Freezing}
%
%
%
\author{Yuanyi~Xue,~\IEEEmembership{Student Member,~IEEE,}
        Beril~Erkin, and~Yao~Wang,~\IEEEmembership{Fellow,~IEEE}
\thanks{Copyright (c) 2014 IEEE. Personal use of this material is permitted. However, permission to use this material for any other purposes must be obtained from the IEEE by sending a request to pubs-permissions@ieee.org.}%
\thanks{Y. Xue, B. Erkin and Y. Wang are with the Department
of Electrical and Computer Engineering, NYU Polytechnic School of Engineering, Brooklyn, NY, 11201 USA. E-mail: \{yxue,be521,yw523\}@nyu.edu.}}
\maketitle
\begin{abstract}
In this work, we propose a novel no-reference (NR) video quality metric that evaluates the impact of frame freezing due to either packet loss or late arrival. Our metric uses a trained neural network acting on features that are chosen to capture the impact of frame freezing on the perceived quality. The considered features include the number of freezes, freeze duration statistics, inter-freeze distance statistics, frame difference before and after the freeze, normal frame difference, and the ratio of them. We use the neural network to find the mapping between features and subjective test scores. We optimize the network structure and the feature selection through a cross validation procedure, using training samples extracted from both VQEG and LIVE video databases. The resulting feature set and network structure yields accurate quality prediction for both the training data containing 54 test videos and a separate testing dataset including 14 videos, with Pearson Correlation Coefficients greater than 0.9 and 0.8 for the training set and the testing set, respectively. Our proposed metric has low complexity and could be utilized in a system with realtime processing constraint.
\end{abstract}


%

\section{Introduction}\label{sec:intro}
\IEEEPARstart{F}{rame} freezing and the consequent temporal jerkiness is a commonly observed artifact in Internet video applications due to both packet losses and delays. Depending on the delay allowance of the underlying video applications, there are two types of frame freezing artifacts. In applications with stringent low delay requirement, (such as video conferencing or live streaming), any frame that is not completely received by its display deadline is considered lost, and the receiver chooses a certain error concealment method to recover the frame. A common and popular error concealment method is to use its previous frame that was correctly received. The subsequent frame, even if correctly received, will have decoding error, if it is predicatively coded using the previous frame. To avoid such error propagation problem, all subsequent frames after a lost frame are also replaced by the same last correctly received frame, until the next intra-frame is received. We call the resulting artifact ``frame freeze due to packet loss''. In applications allowing more elastic delay, such as streaming of pre-coded video, when a frame does not arrive past its display deadline, the receiver continuously displays the previous frame, until the actual new frame arrives. We call this artifact ``frame freeze due to packet delay''. Both artifacts manifest as temporal jerkiness on the received video. In this work, we propose a metric that quantitatively measures this temporal jerkiness and its relation to the subjective quality.

The proposed metric is a no-reference (NR) metric, in that it evaluates the jerkiness based on the degraded video only, rather than by comparing the degraded video with the original pristine video. No-reference metrics are important for quality assessment in real applications, as the pristine video is often not available at the receiver. There are several related works for detecting and measuring the frame freeze. In~\cite{Quan_ICIP09_NRfreeze}, the authors proposed an NR metric for measuring the quality impact of frame freeze, based on the profiles of the duration of each freeze and the number of freezes. However, the features they considered are not dependent on the video content, which is undesirable, since for different video characteristics, the same freeze pattern could have different impacts on the quality. In~\cite{NTIA_freezeDetect}, the authors proposed an algorithm for detecting the freeze frames. The algorithm uses the absolute and squared values of 1-step frame differences, and detects the possible dropped frames by comparing to thresholds that depend on the inherent motion characteristics of the video. In a more recent work~\cite{Yammine_FreezeDetec_PCS2012}, the authors also used the squared value of 1-step frame differences, but added an extra encoding pass for the received video. Based on the fact that the frames in a freeze event are exactly the same, the authors' detection algorithm used different thresholds according to the frame types of neighboring frames after that additional encoding. Even though the method of~\cite{Yammine_FreezeDetec_PCS2012} provides more accurate freeze detection than that reported in~\cite{Quan_ICIP09_NRfreeze}, the required re-encoding process in the method will bring too much overhead to be deployed in a real-time system. In~\cite{TemporalMetric_2013}, the authors measured the number of freezes as well as the average freeze duration in each preset segment (e.g., the first few seconds, the middle part and the last few seconds) and characterize the quality of each segment by the product of the number of freezes and the average freeze duration. Finally the overall quality of the entire video is modeled using an exponential mapping of a weighted sum of qualities of each segment. The weights depend on the positions of the segments in the video. This metric considers the fact that the degradation caused by frame freeze depends on the freeze locations, but it is still content-independent. In the recent NR metric standardized by ITU-T, namely Rec. ITU-T P.1201.1/.2~\cite{itu1201.1, itu1201.2, VPQM_1201}, which relies on packet header information, the frame freezing quality degradation is estimated by the ratio of the number of damaged video frames and the total number of video frames as well as the packet loss event frequency. Again, the method did not consider the content characteristic. Furthermore, it does not differentiate between random individual frame drops and consecutive frame losses. Finally this metric involves a series of complicated mapping functions, making it difficult to compute.

With the shortcomings of previous works in mind, our proposed algorithm operates on the decoded video (thus no header information is needed) and it explicitly considers the differences in video content by measuring frame difference statistics. Furthermore, instead of trying to find an analytical mapping between features and the perceived quality, we use a neural network to learn the mapping from the training data. The remainder of the paper is organized as follows: in Section~\ref{sec:algo} we will introduce our algorithm for freeze detection and feature extraction. In Section~\ref{sec:model} we report our model selection and training procedure. We show our result in Section~\ref{sec:res}, and conclude the paper in Section~\ref{sec:conc}.
\section{Proposed algorithm}\label{sec:algo}
\begin{figure}[!t]
\centering
\includegraphics[width=0.85\columnwidth]{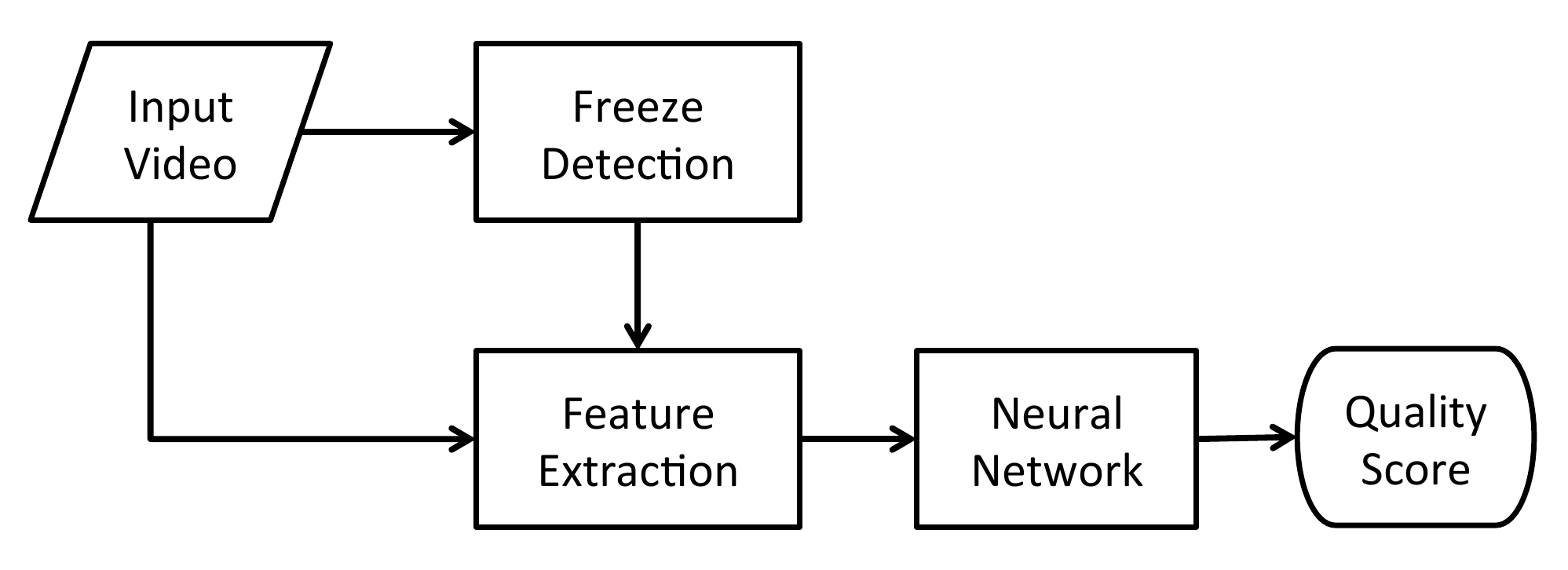}
\caption{Flowchart of the proposed metric.}
\label{fig:flowchart}
\vspace{-.2in}
\end{figure}

Our proposed algorithm consists of three main parts, including detecting freeze frames, extracting features and mapping the features into a quality score. We describe each step in the following sections. The overall flowchart of the proposed metric is illustrated in Fig.~\ref{fig:flowchart}.
\subsection{Freeze frame detection}
Freeze frame detection is a crucial step for our quality metric. A na\"{\i}ve way to finding freeze frames is to examine the difference between each frame and its previous frame, and label the frames that have zero frame difference as freeze frames. This is based on the nature of a freeze event, as the exact same frame is held during the entire freeze duration. In Fig.~\ref{fig:freezeprof} and \ref{fig:delayd}, we show frame difference profiles from two videos used in this study, where the frame difference ($\mathcal{FD}$) is defined below:
\begin{equation}
\mathcal{FD}(i) = \mathbf{MEAN}((\mathbf{Y}(i+1) - \mathbf{Y}(i))^2)
\label{eq:TI2}
\end{equation}
in which $\mathbf{Y}$ is the Y-channel image of a frame. Both Fig.~\ref{fig:freezeprof} and Fig.~\ref{fig:delayd} have four segments during which the frame difference is zero. They correspond to four freeze events.

Although the zero frame difference appears to be reasonable for detecting frame freezes, in practice, as the quality metric is often computed from the captured video frames rendered on the display screen, there could be small differences between displayed duplicated frames. Therefore, a more robust way to detecting frame freeze is by comparing the frame difference to a non-zero threshold. The challenge lies in how to set the threshold. For example, in the case of videos with large portion of static scenes, even a very small threshold may yield some false alarms. In this work, we use the freeze detection algorithm proposed in~\cite{NTIA_freezeDetect}. We use the same set of thresholds suggested in~\cite{NTIA_freezeDetect} for both packet loss and packet delay videos.

In Table~\ref{tab:freezecnt}, we report the true freeze detection rate, and false alarm rate for both VQEG and LIVE videos we use in this work. The freeze detection algorithm performs accurately for LIVE videos, while it has some difficulties in picking up freezes for two VQEG videos containing mostly low and smooth motions. We would like to note that the same difficulty would be also applicable to human subjects when the video consists of low and smooth motion. And because of the ambiguity of the freeze and smooth motion in such cases, it is therefore expected that the missed freeze events will not lead to perceivable quality degradation in perceptual quality. We would also like to note that the algorithm will discard any detected freeze event containing less than 2 frames, which is usually not perceivable by the human observer.
\begin{table}[!t]
\centering
\caption{Detection and false alarm of freeze detection algorithm}
\label{tab:freezecnt}
\small{
\begin{tabular}{|c||c|c|c|c|c|}
\hline
 &  & Correctly &  &  &  \\ 
 & Total & detected & Detection & False & False \\
Dataset & freezes & freezes & rate & alarms & alarm rate \\ \hline \hline
VQEG & $140$ & $131$ & $93.57\%$ & $11$ & $7.75\%$ \\ \hline
LIVE & $160$ & $160$ & $100.00\%$ & $12$ & $6.89\%$ \\ \hline
\end{tabular}
}
\vspace{-.1in}
\end{table}
\subsection{Feature extraction}\label{sec:featext}
After we get the freeze frame locations for a giving video, we extract a total number of 13 features for a video.
\begin{table*}[!t]
\centering
\small{
\caption{Summary of 13 features}
\label{tab:featlist}
\begin{tabular}{|l||l|}
\hline
Feature name & Description \\ \hline \hline
\textbf{NumFz} & Number of freeze events \\ \hline
\textbf{AvgFzDur, MaxFzDur, StdFzDur} & Average, maximum and standard deviation of freeze durations \\ \hline
\textbf{AvgFzDist, MaxFzDist, StdFzDist} & Average, maximum and standard deviation of distances in between two freezes \\ \hline
\textbf{rLenFz} & Ratio of total freeze length vs. length of the video \\ \hline
\textbf{rDurDist} & Ratio of average freeze duration and the average inter-freeze distance \\ \hline
\textbf{AvgFzFD, MaxFzFD} & Average and maximum post freeze frame difference\\ \hline
\textbf{AvgBgFD} & Average frame difference of video content excluding freezes and scene cuts \\ \hline
\textbf{rFD} & Ratio of \textbf{AvgFzFD} and \textbf{AvgBgFD} \\ \hline
\end{tabular}
}
\vspace{-.1in}
\end{table*}

First of all, we identify each consecutive set of freeze frames as a freeze event and calculate the duration (in terms of number of frames) of each freeze event, and the distance (also in terms of number of frames) between every two adjacent freeze events. The number of freezes (\textbf{NumFz}) is the number of freeze events for the whole video. \textbf{AvgFzDur}, \textbf{MaxFzDur}, and \textbf{StdFzDur} are the mean, maximum, and standard deviation of the freeze durations among all freeze events. \textbf{AvgFzDist}, \textbf{MaxFzDist}, and \textbf{StdFzDist} are the mean, maximum, and standard deviation of the distances in between two freezes among all freeze events. Finally, \textbf{rLenFz} is the ratio of the total freeze length versus the length of the video, and \textbf{rDurDist} is the ratio of the average freeze duration and the average inter-freeze distance.

\begin{figure}[!t]
\centering
\includegraphics[width=0.8\columnwidth]{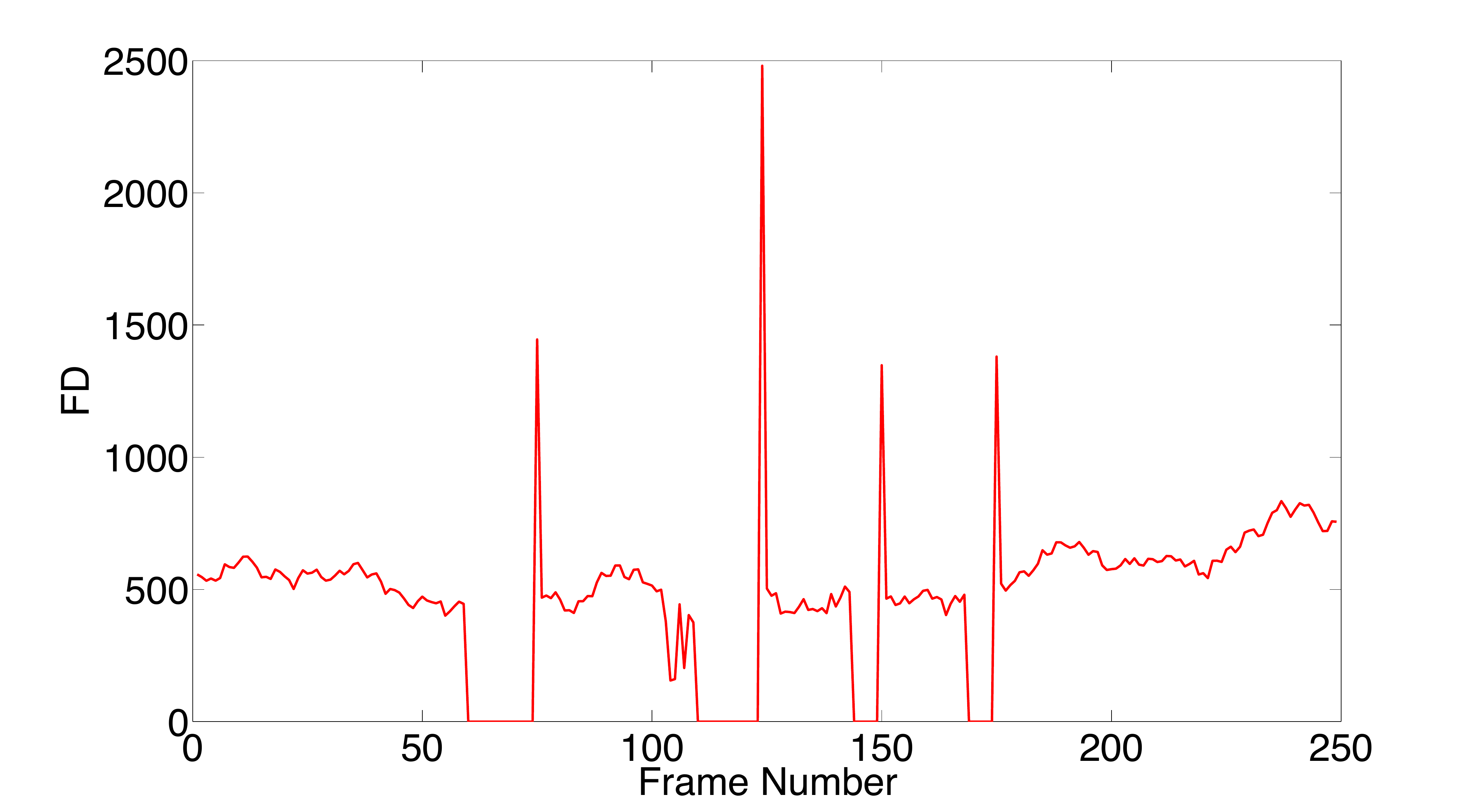}
\caption{Frame difference profile for a video suffering from frame freeze due to frame loss, Y channel.}
\label{fig:freezeprof}
\end{figure}

The aforementioned features only depend on the freeze event pattern, but not the actual video content. We also explicitly construct features that are related to the video content, more specifically the frame difference information, which depends on the motion characteristics. The first two features are used to describe a phenomenon typically observed in frame freeze due to packet loss. As shown in Fig.~\ref{fig:freezeprof}, we can see that there are $\mathcal{FD}$ peaks immediately after a freeze due to packet loss. In this paper, we refer this as the ``freeze frame difference'', denoted by \textbf{FzFD}. Intuitively, the frame difference after a freeze is proportional to the length of the freeze, as well as the actual motion during the freeze. Therefore, we define two content-dependent features, denoted by \textbf{AvgFzFD} and \textbf{MaxFzFD}, which are the mean and maximum of such \textbf{FzFD} among all the freeze events for a video.

To compute the background frame difference, we take the average of frame differences over all frames, excluding both the freeze region and scene cut region. In this work, scene cut detection is based on the following heuristic rule: we consider a frame is during a scene cut if the frame difference of the current frame is larger than 5 times the average frame difference of previous five frames.

\begin{figure}[!t]
\centering
\includegraphics[width=0.8\columnwidth]{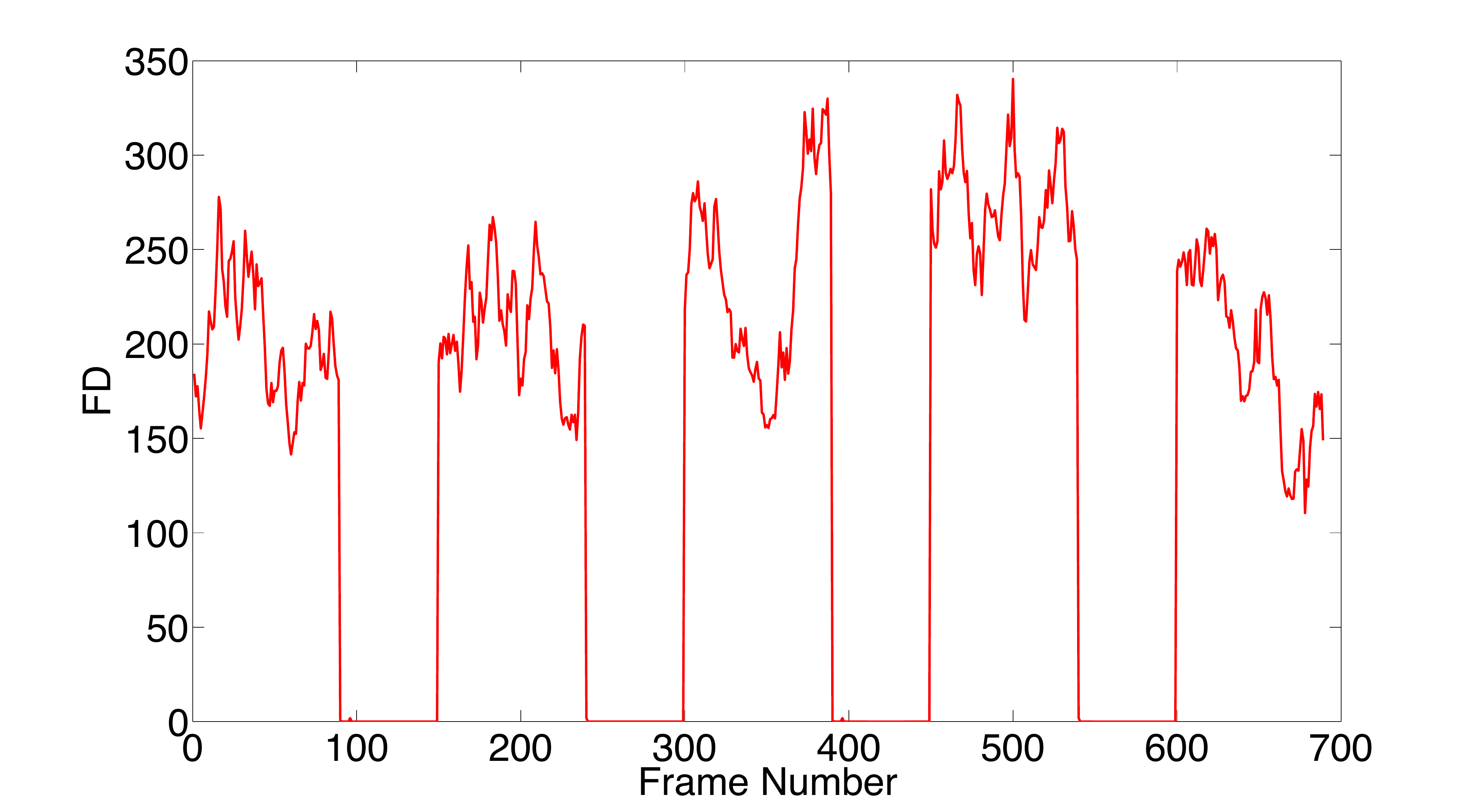}
\caption{Frame difference profile for a video suffering from frame freeze due to packet delay, Y channel.}
\label{fig:delayd}
\end{figure}

We name this feature as the background frame difference (\textbf{AvgBgFD}). Lastly, the ratio of \textbf{AvgFzFD} and \textbf{AvgBgFD} is used as another feature (\textbf{rFD}). Note that for frame freeze due to packet delay, the frame difference immediately after a freeze will be similar to background frame difference, as illustrated in Fig.~\ref{fig:delayd}. Therefore, these features are expected to be less useful for characterizing impact of frame freezes due to delay.
In Table~\ref{tab:featlist} we give a summary of 13 features considered in developing the proposed metric. We would like to note that in a streaming scenario like videoconference, the features can be estimated using the received video. Na\"{\i}vely, for features involving counting, one could increment the count whenever an event (e.g. freeze) happens; for features related to content, one could use all the received frames up to the last known freeze location to derive the features as the input of the proposed metric. And the received frames can be used to update those estimates periodically or after a freeze completes. The engineering detail of how to implement the proposed metric to real video streaming system, is nevertheless beyond the scope of this paper.
\section{Model selection and training}\label{sec:model}
\subsection{Dataset}
We use two publicly available annotated video databases as our dataset. The VQEGHD dataset 5~\cite{VQEG_video} contains a total of 28 processed video sequences (PVSs) with frame freezes due to packet loss. The PVS configuration is listed in Table~\ref{tab:vqeg}.
\begin{table}[!t]
\centering
\small{
\caption{PVS configuration for VQEGHD5 database}
\label{tab:vqeg}
\begin{tabular}{|c||c|}
\hline
Video contents & 7 source videos, 28 PVSs\\ \hline
Spatial resolution & 1080p \\ \hline
Temporal resolution & 25Hz \\ \hline
Duration & 10 seconds \\ \hline
Encoder & H.264 \\ \hline
Bitrate & 16Mpbs \\ \hline
PLR & Bursty 0.125, 0.25, 0.5, 1 and 2\\ \hline
Freeze type & Freeze due to packet loss \\ \hline
\end{tabular}
}
\vspace{-.1in}
\end{table}

\begin{figure*}[!t]
\centering{
\subfloat[]{\includegraphics[scale=0.05]{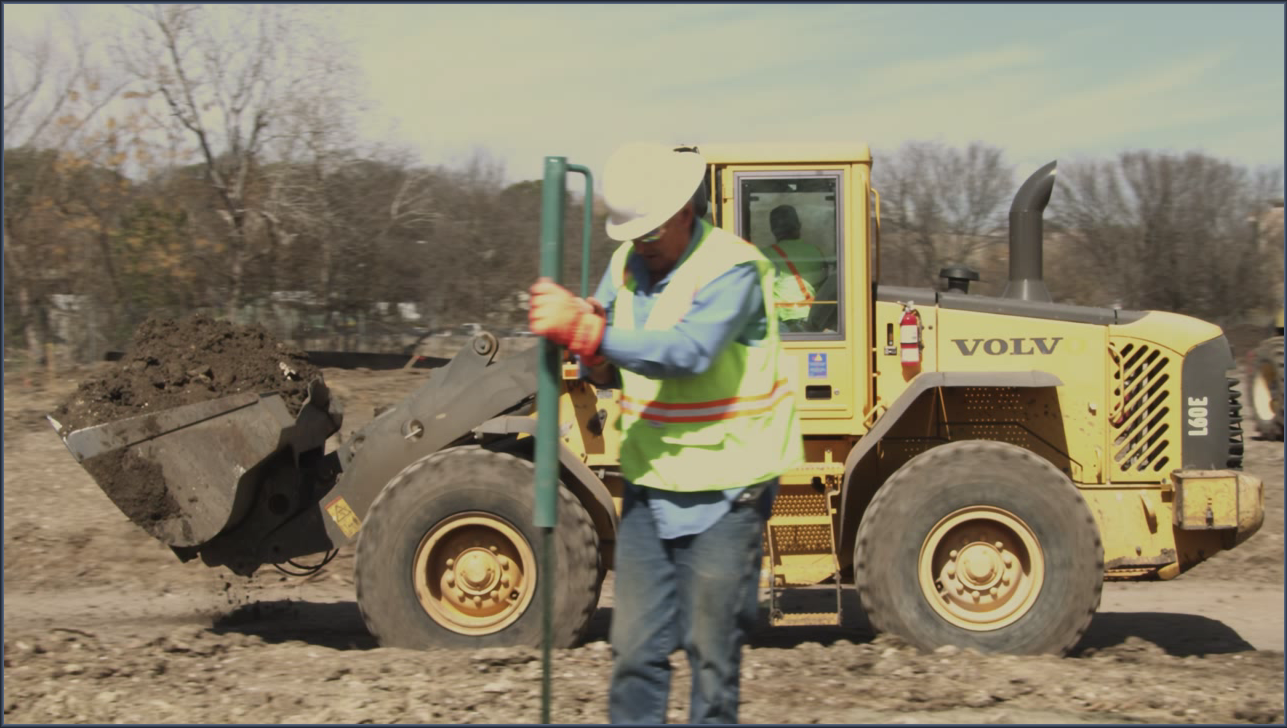}}
\subfloat[]{\includegraphics[scale=0.05]{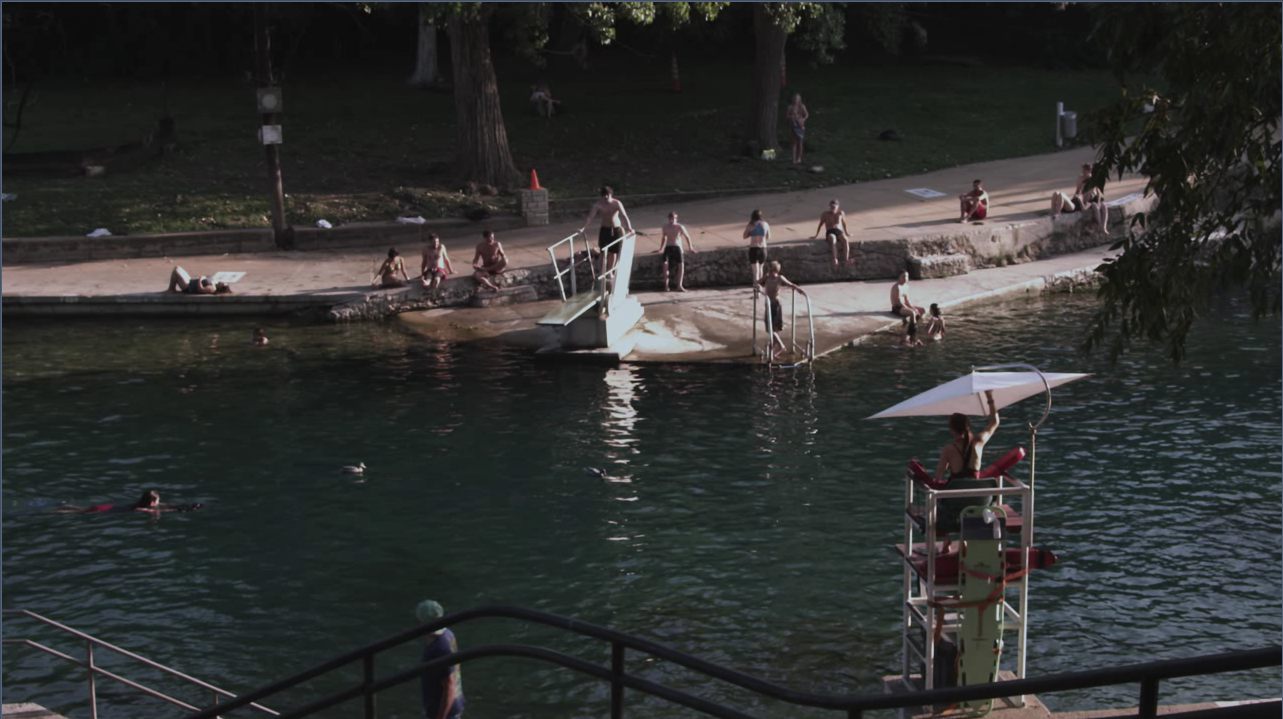}}
\subfloat[]{\includegraphics[scale=0.05]{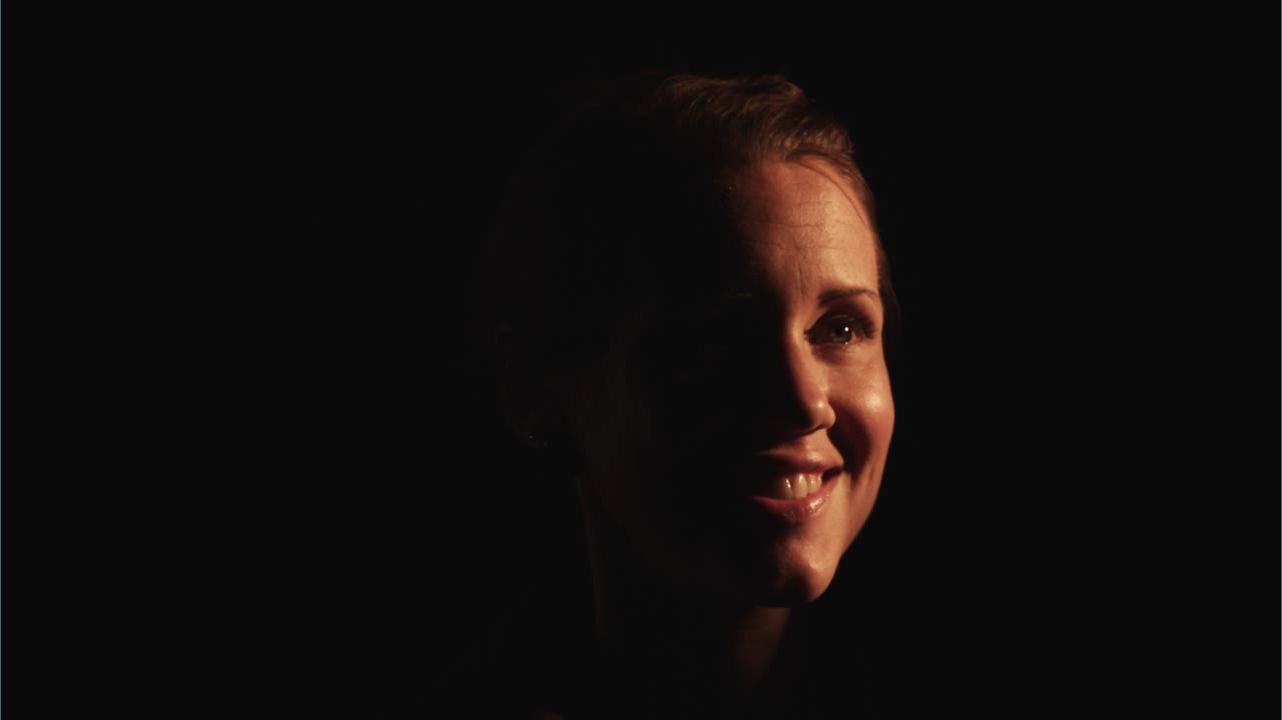}}
\subfloat[]{\includegraphics[scale=0.05]{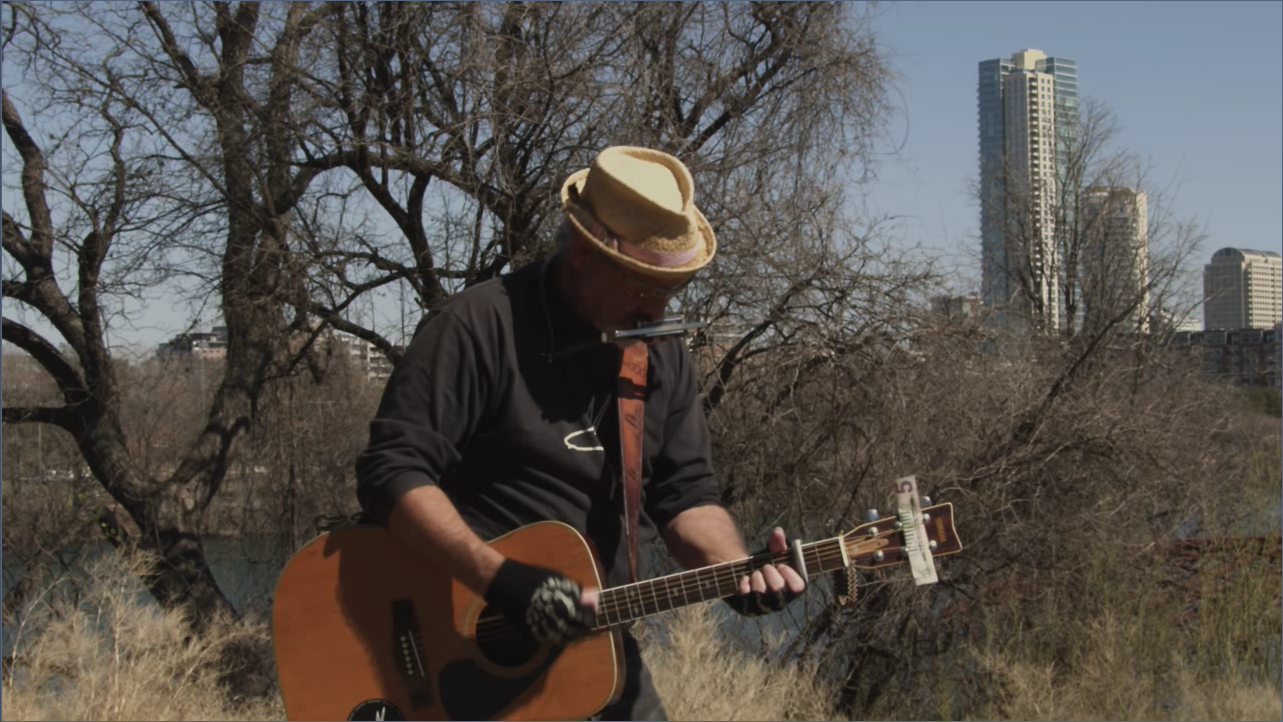}}
\subfloat[]{\includegraphics[scale=0.05]{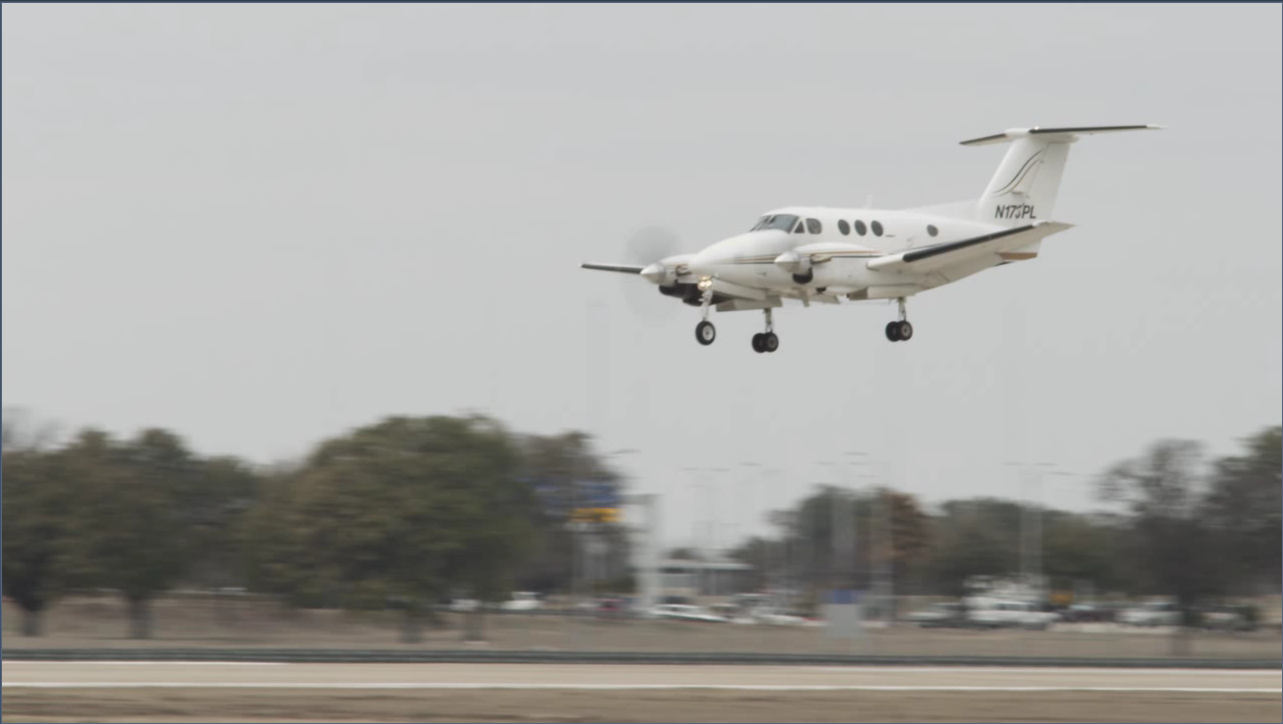}}
\subfloat[]{\includegraphics[scale=0.05]{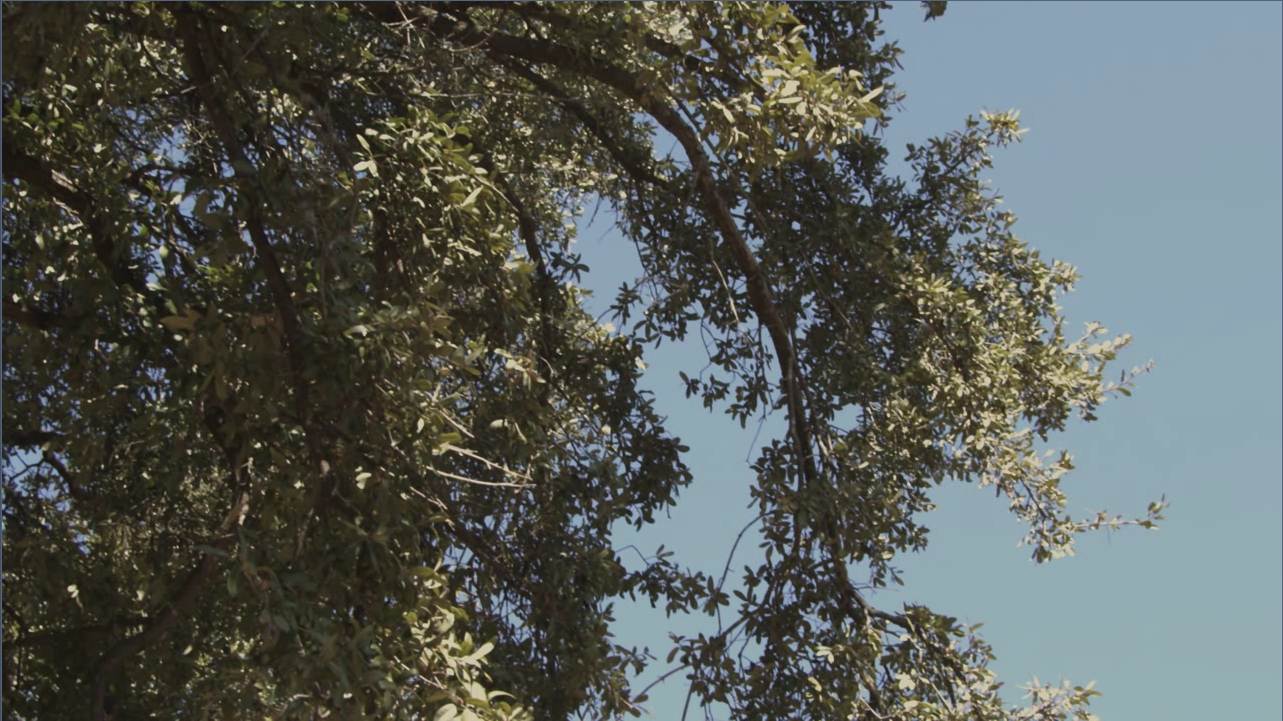}}
\subfloat[]{\includegraphics[scale=0.05]{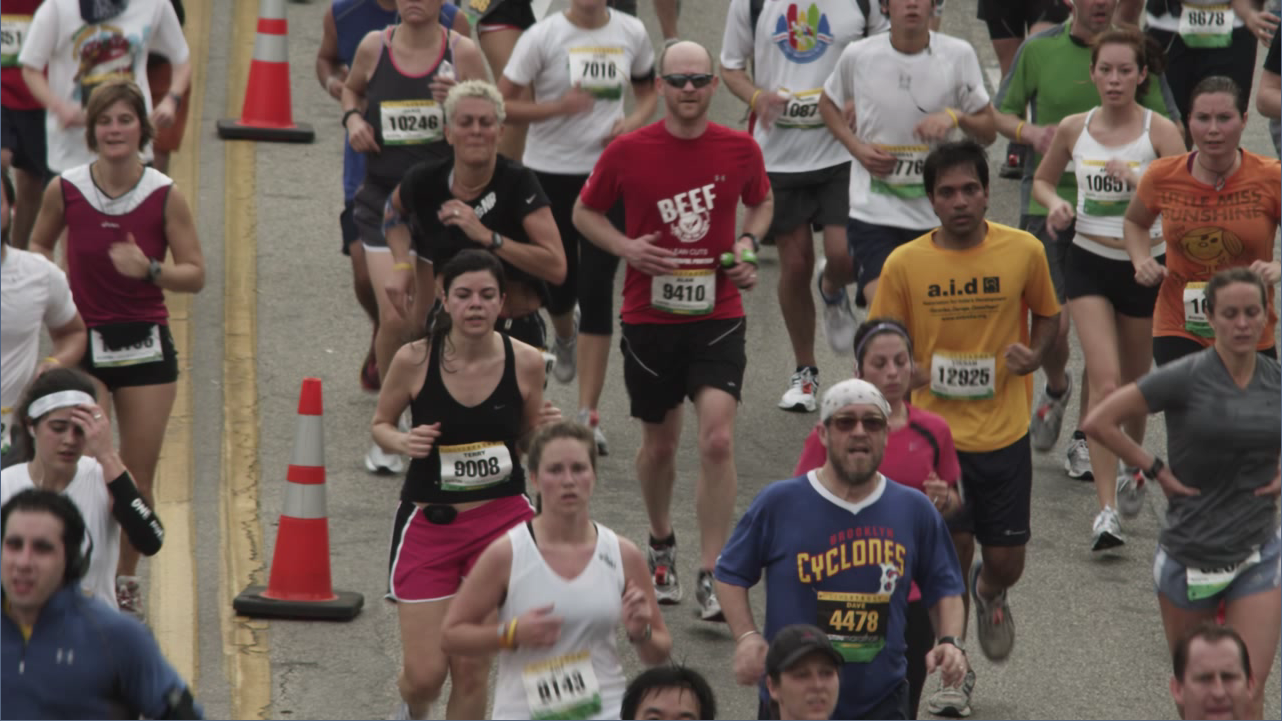}}
\subfloat[]{\includegraphics[scale=0.05]{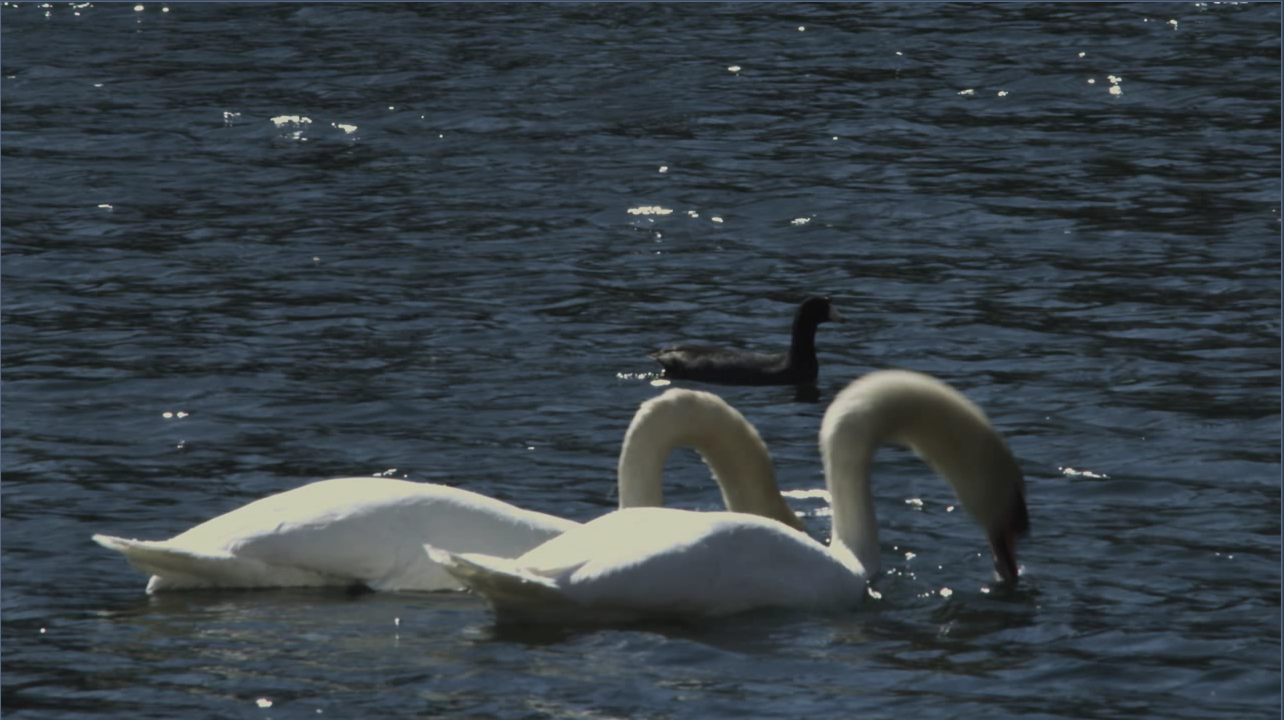}}\\
\subfloat[]{\includegraphics[scale=0.05]{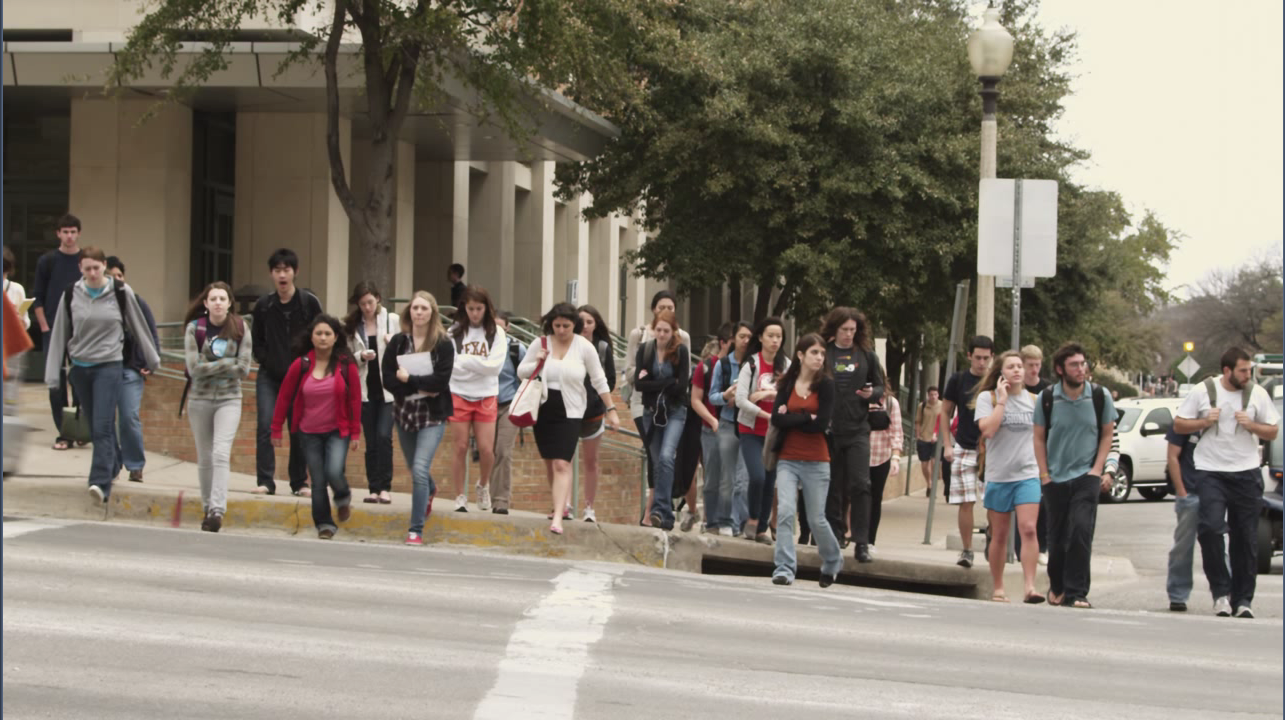}}
\subfloat[]{\includegraphics[scale=0.05]{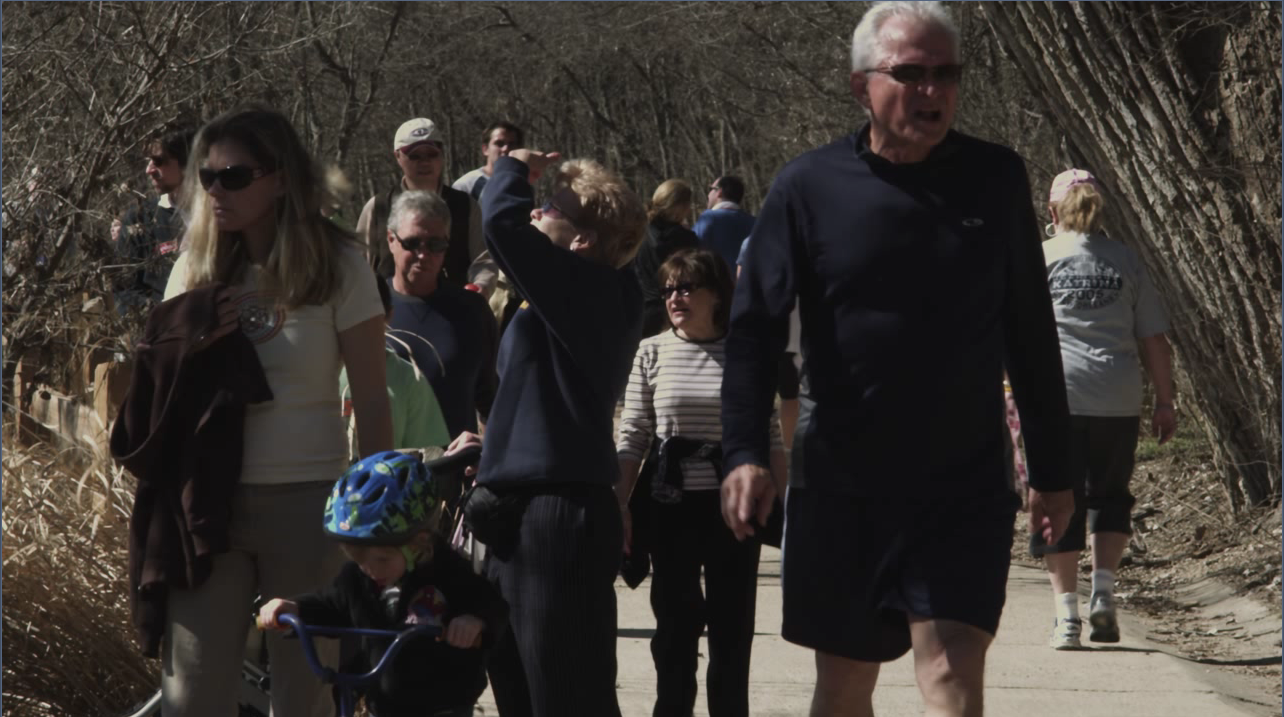}}
\subfloat[]{\includegraphics[scale=0.03]{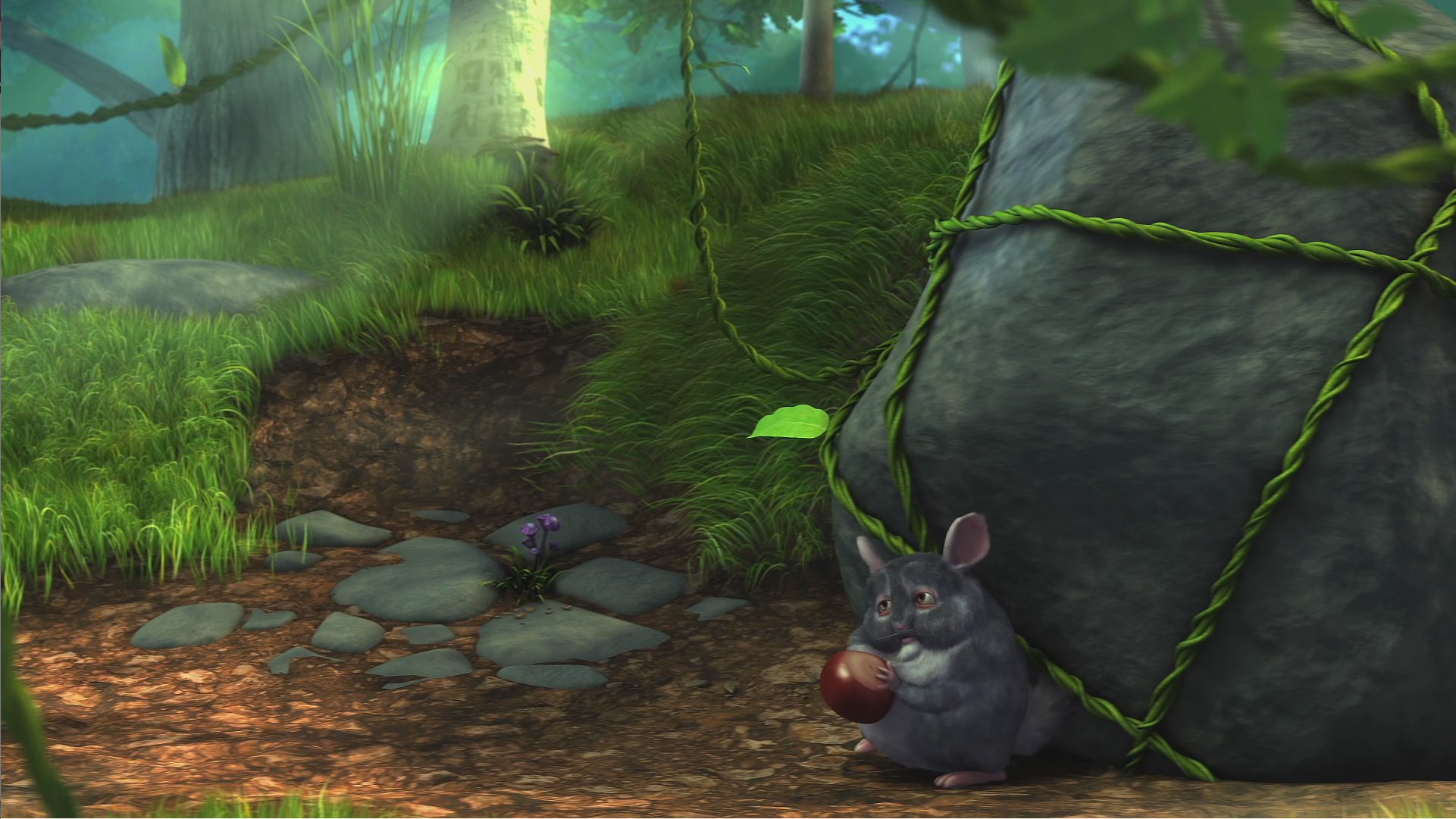}}
\subfloat[]{\includegraphics[scale=0.03]{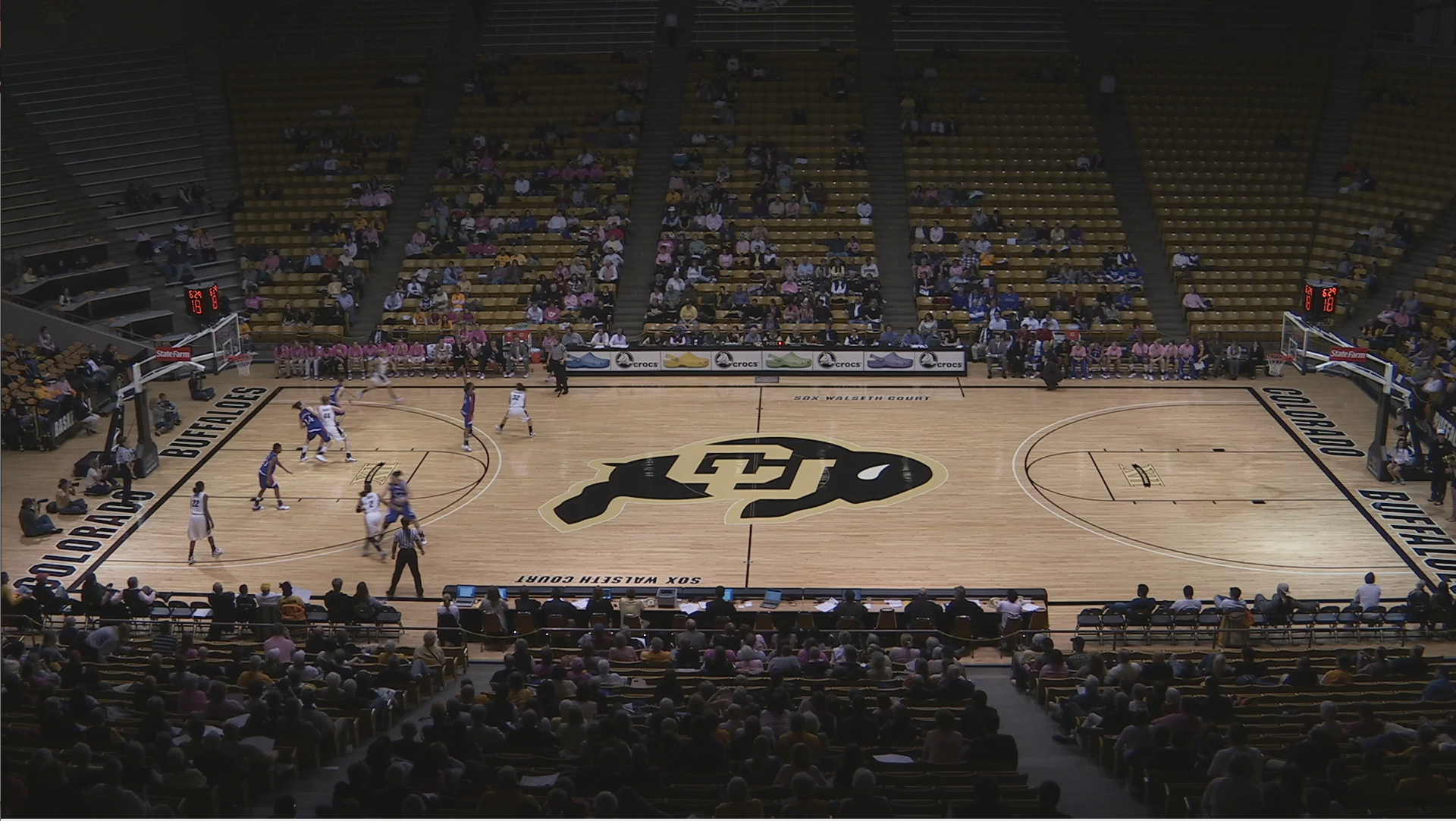}}
\subfloat[]{\includegraphics[scale=0.03]{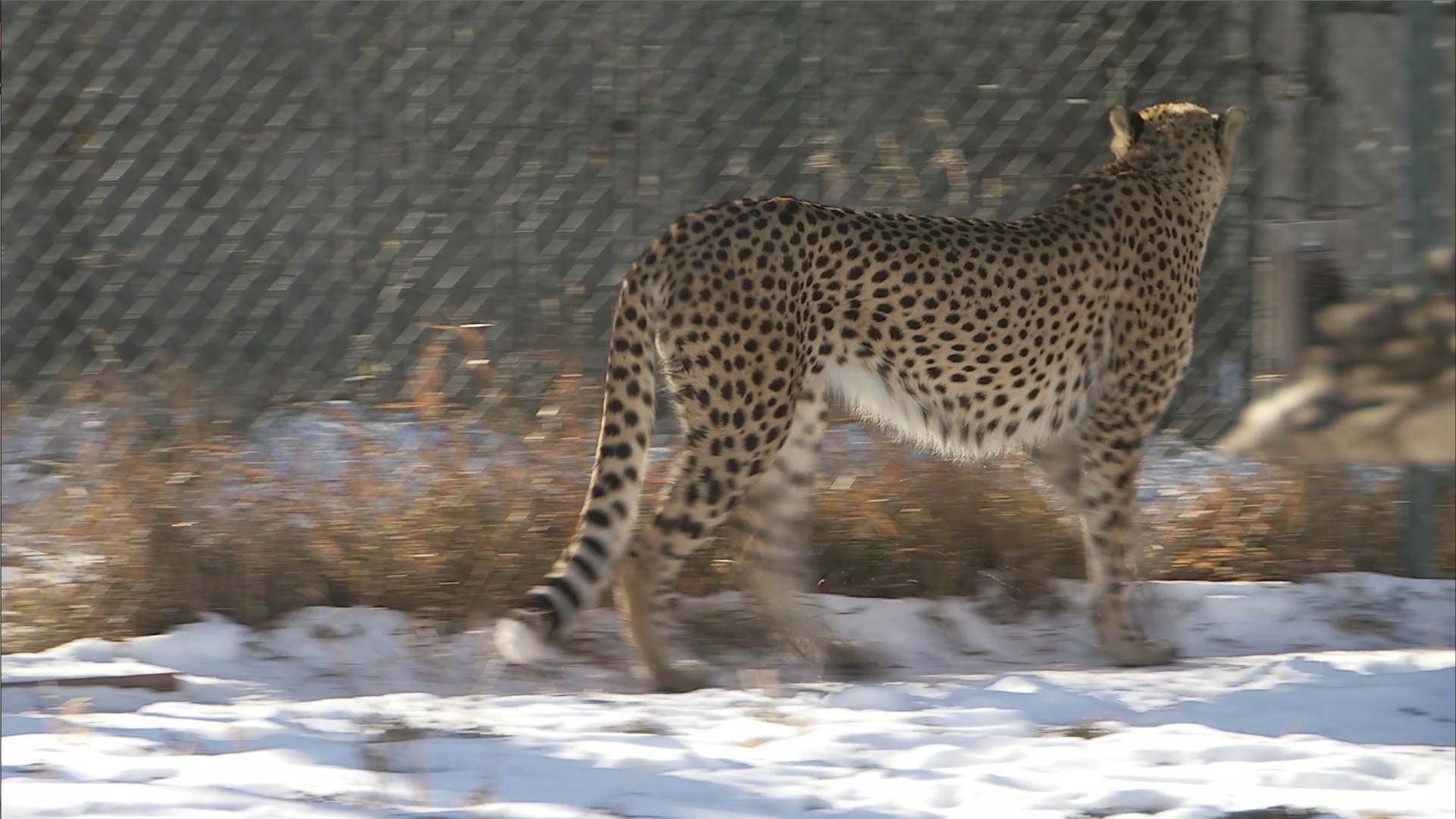}}
\subfloat[]{\includegraphics[scale=0.03]{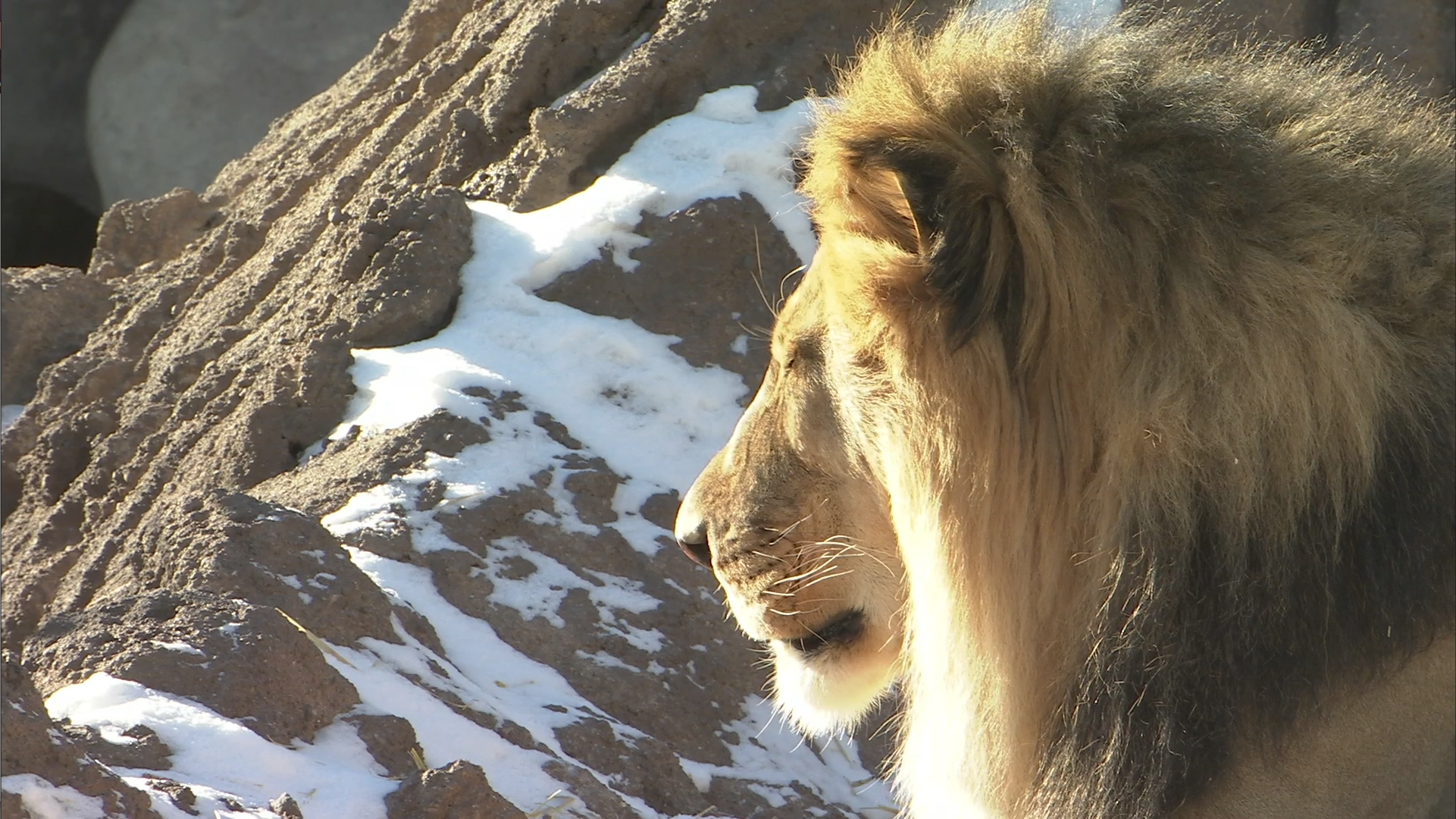}}
\subfloat[]{\includegraphics[scale=0.03]{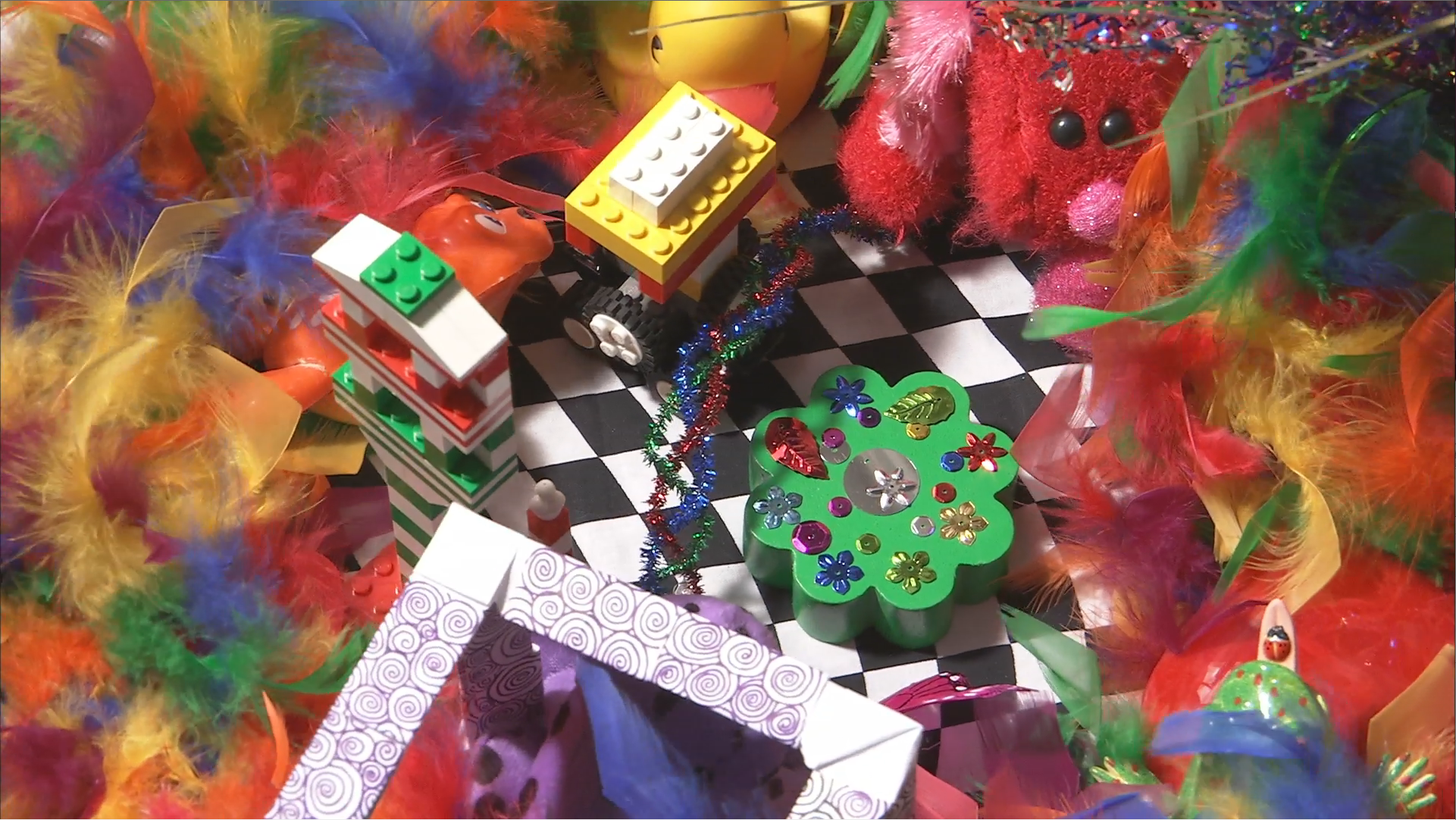}}
\subfloat[]{\includegraphics[scale=0.03]{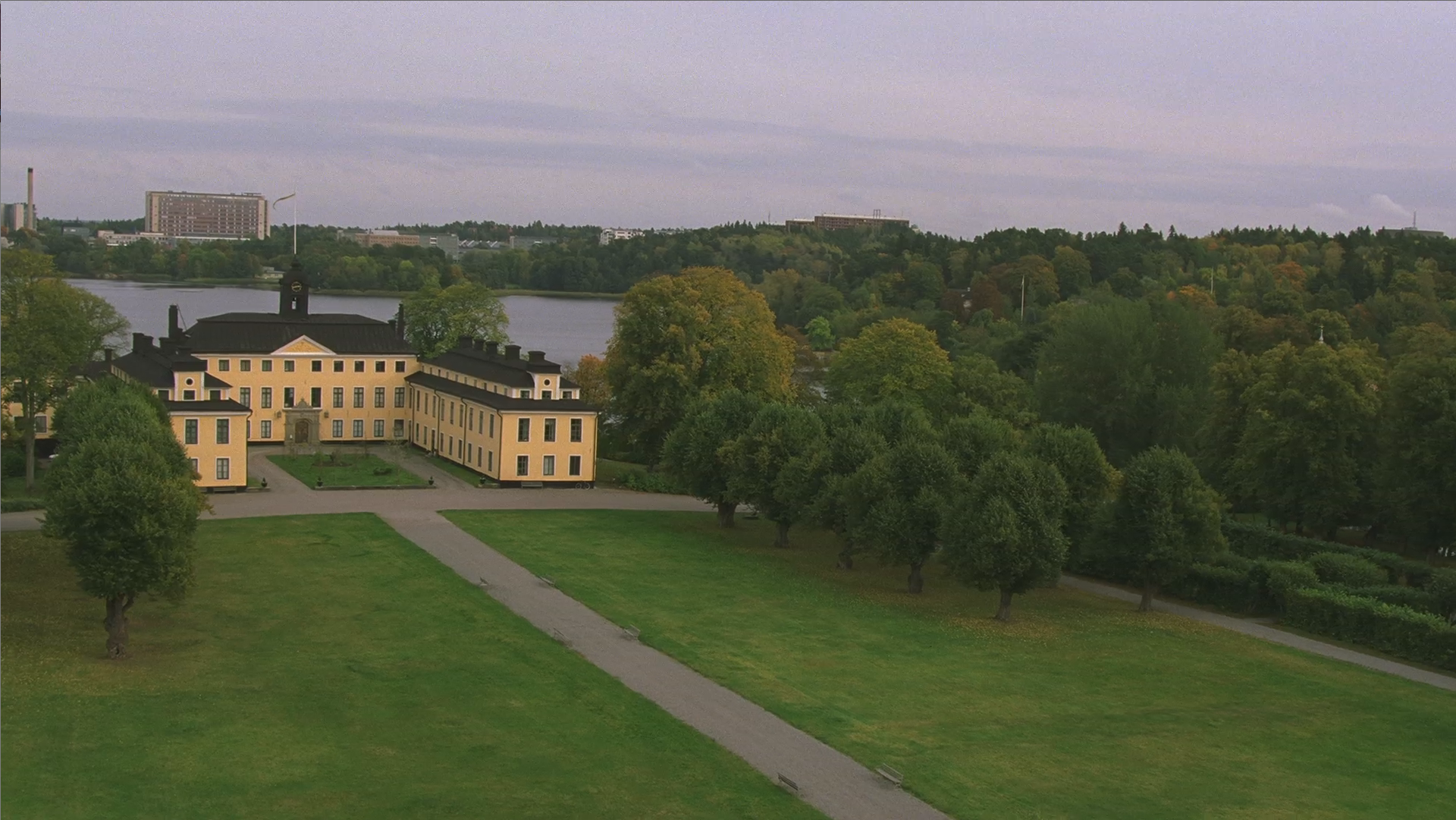}}
\subfloat[]{\includegraphics[scale=0.03]{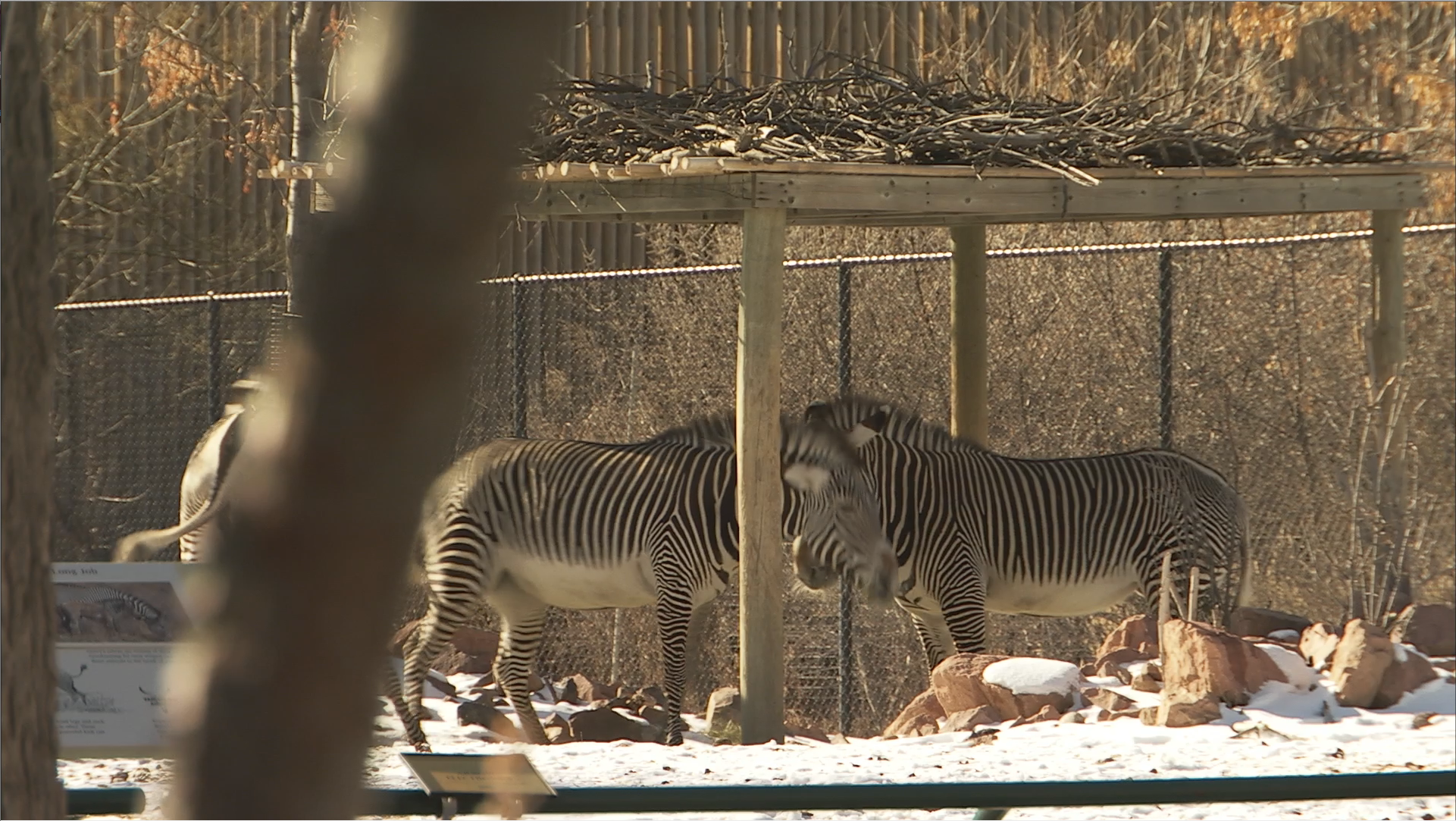}}
\caption{Snapshots for videos used in this work}
\label{fig:pvs}}
\end{figure*}

\begin{figure}[!t]
\centering
\includegraphics[width=0.75\columnwidth]{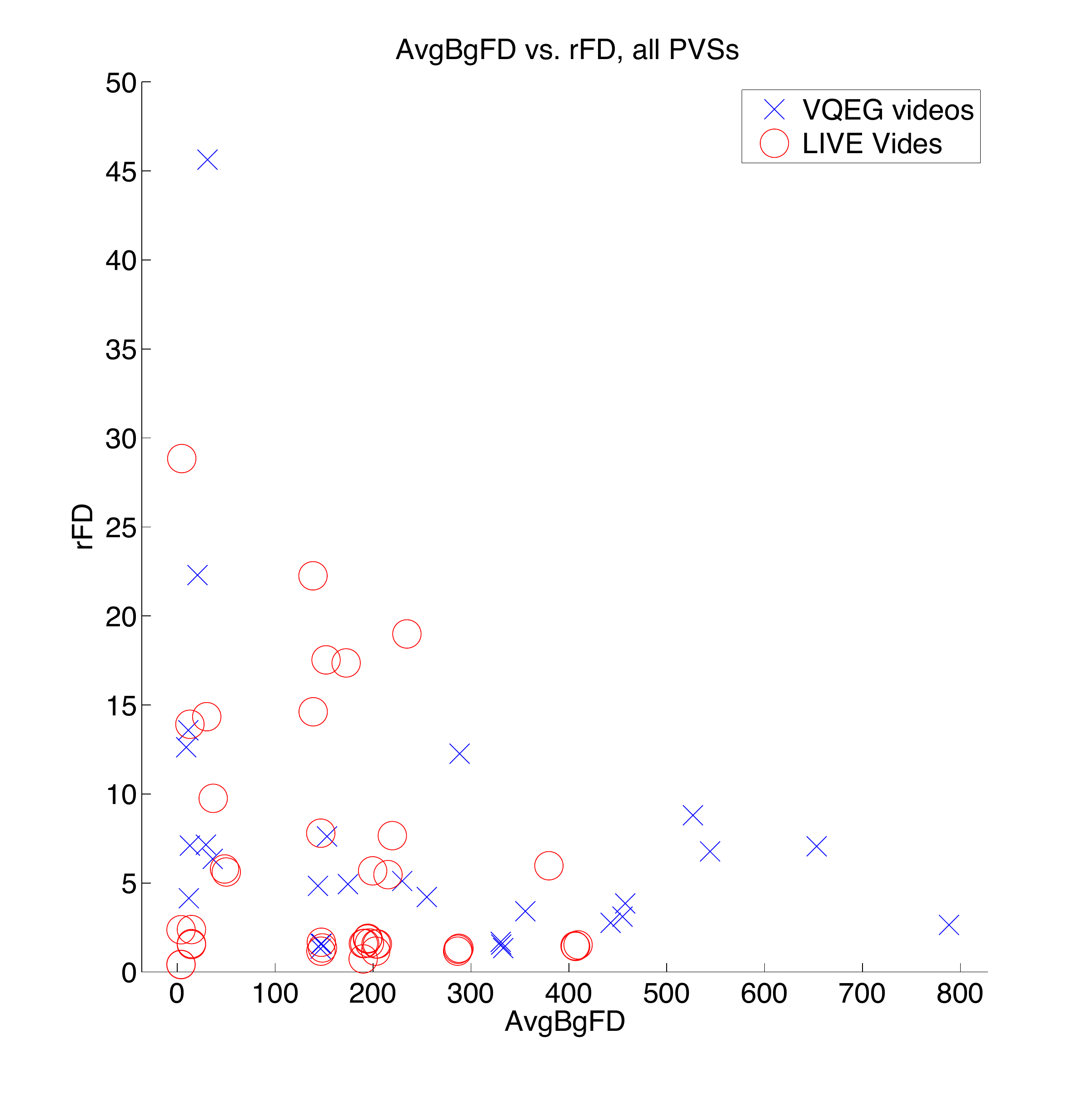}
\caption{Scatter plot of AvgBgFD vs. rFD.}
\label{fig:motionscatter}
\end{figure}

The LIVE mobile FF database~\cite{LIVE_database} contains a total of 40 PVSs, 30 of them have frame freezes due to the packet delay (which have no observed 1-step frame difference spikes immediately after a freeze event, see Fig.~\ref{fig:delayd} for reference), while 10 of them have frame freezes due to packet loss. The PVS configuration for LIVE database is listed in Table~\ref{tab:live}. We show the screenshots of source videos in Fig.~\ref{fig:pvs}. In Fig.~\ref{fig:motionscatter}, we also show a scatter plot of AvgBgFD vs. rFD of all 68 PVs in our database. The scatter plot shows that the considered PVSs cover a large variety of different motion characteristics.
\begin{table}[!h]
\centering
\small{
\caption{PVS configuration for LIVE database}
\label{tab:live}
\begin{tabular}{|c||c|}
\hline
Video contents & 10 source videos, 40 PVSs \\ \hline
Spatial resolution & 720p \\ \hline
Temporal resolution & 30Hz \\ \hline
Duration & 15 seconds \\ \hline
Encoder & RAW \\ \hline
PLR & Every 1, 2 and 4 seconds\\ \hline
Freeze Type & 30 PVSs due to packet delay, \\
& 10 PVSs due to packet loss. \\ \hline
\end{tabular}}
\vspace{-.1in}
\end{table}

We split the whole dataset into a training set and a testing set with $80/20$ proportion, resulting a training set with 52 PVSs from 13 source videos and a testing set with 16 PVSs from 4 source videos. The training set is used to optimize the feature selection and neural network structure and weights. The chosen feature set and network structure and the trained neural network weights are then applied to the testing set to evaluate the performance of the proposed metric. Our testing set only includes source videos not used in the training set. In the testing set, we have a total of 4 source videos (16 PVSs) covering different levels of texture and motion activities. The corresponding snapshots of videos in the testing set are shown in Fig.~\ref{fig:pvs}(d)(e)(m)(p). The source videos in both the training set and testing set, respectively, cover typical motion and texture characteristics.

We use the difference mean opinion score (DMOS) as the subjective quality measure for our dataset. The DMOS score is defined as $MOS(PVS)-MOS(SRC)+5$, where SRC is the hidden reference source.
\subsection{Neural Network Structure Selection}
\begin{figure}[!t]
\centering
\includegraphics[width=0.7\columnwidth]{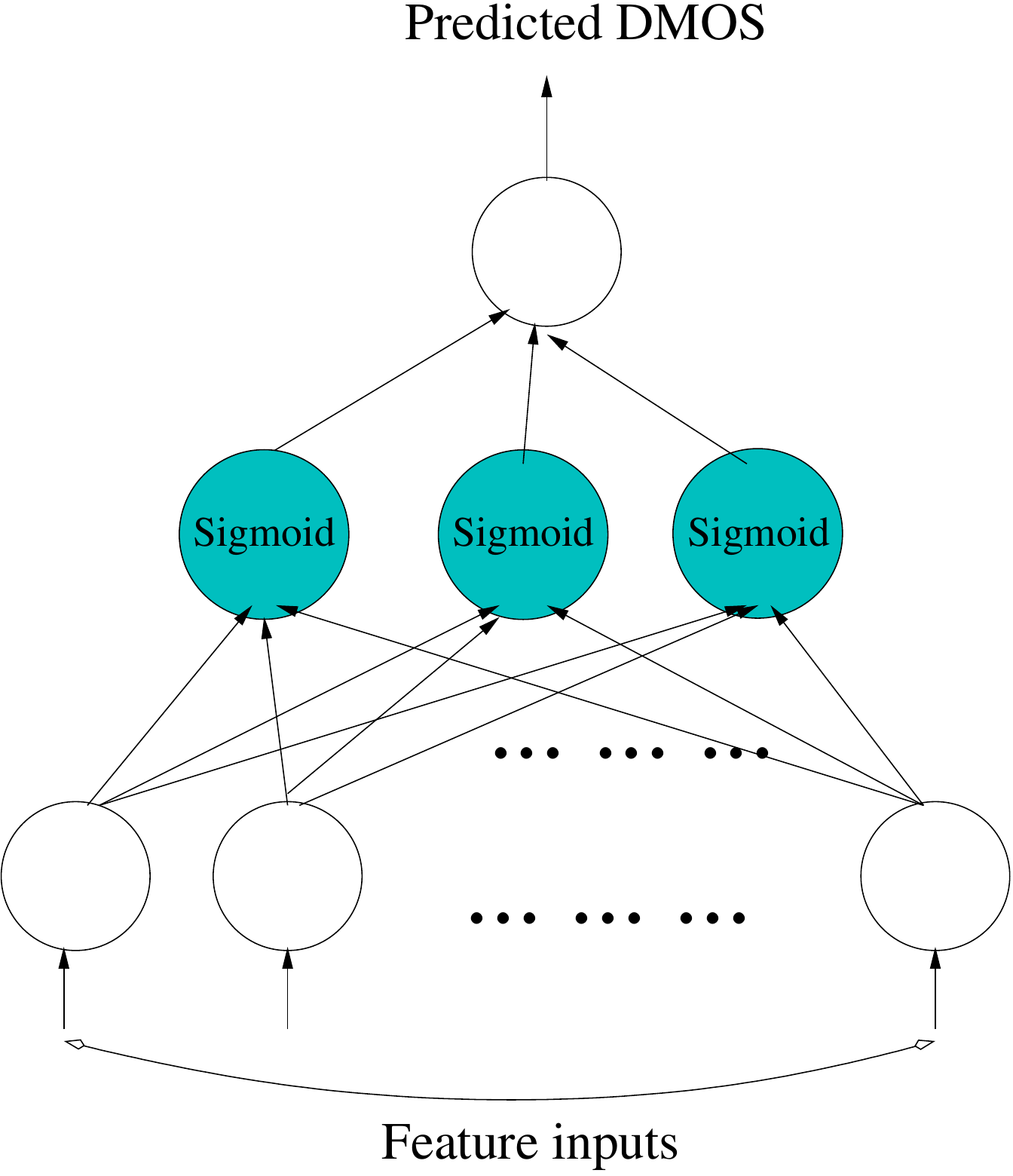}
\caption{Neural network architecture, the hidden layer neurons shown in cyan are with sigmoid transfer function.}
\label{fig:nn}
\vspace{-.1in}
\end{figure}

In this work, instead of using a preset function to find the mapping from the extracted features to the quality score, we use neural network based approach. Considering the limited number of training samples, we only consider a neural network with 1 hidden layer with Sigmoid transfer function. Besides the hidden layer neurons, all other neurons (both output neuron and input neurons) are set to be linear. We ran through different number of hidden layer neurons and found that three nodes in the hidden layer gives the best performance (in a scheme described in the following section). Figure~\ref{fig:nn} shows the network diagram.
\subsection{Feature selection and neural network parameter optimization}
The one layer neural network can generally accommodate any number of input features, and can have any number of hidden nodes. When the number of input features is $\mathcal{N}$ and the number of hidden nodes is $\mathcal{M}$, the total number of weights to be trained is $\mathcal{M}\times (\mathcal{N}+1)+\mathcal{M}+1$. Given the limited number of training samples (we have a total of 52 PVSs in training set), we must choose $\mathcal{M}$ and $\mathcal{N}$ so that $\mathcal{M}\times (\mathcal{N}+1)+\mathcal{M}+1$ is strictly less than 52, to avoid overfitting, and that the $\mathcal{N}$ features should be chosen from all 13 features described in Sec.~\ref{sec:featext} to maximize a proper performance criterion. We combine our feature selection with the model selection (i.e., $\mathcal{M}$) using the popular 10-fold cross validation procedure~\cite{stats_book}, that splits the training set into 10 equally numbered batches, each time one batch is used as the validation fold while the rest are used to train the specific network structure of interest. The training procedure is repeated 10 times and the performance is reported as the average error on the validation set.

We use the exhaustive search for selecting both the number of hidden nodes $\mathcal{M}$ and the number of features $\mathcal{N}$. Furthermore, for a given $\mathcal{N}$, we search through all possible feature combinations. We find the combination of $\mathcal{M}$ and $\mathcal{N}$ and corresponding $\mathcal{N}$ features that leads to the minimal cross validation error. Through this procedure we find that $\mathcal{M}=3$ hidden nodes and $\mathcal{N}=6$ features give the best performance. Table~\ref{tab:feats} lists the chosen six features.
\begin{table}[!t]
\centering
\small{
\caption{Chosen Features Through Exhaustive Search}
\label{tab:feats}
\begin{tabular}{|c||l|}
\hline
Features & AvgFzDist, NumFz, rDurDist, rFD, StdFzDist, rLenFz \\ \hline
\end{tabular}}
\vspace{-.1in}
\end{table}
\subsection{Training and testing}
For given $\mathcal{N}$ input features and $\mathcal{M}$ hidden nodes, and for each of the 10 possible training subsets (in the 10-fold cross validation process), we follow the conventional method for training the network weights, to minimize the mean square error between the predicted quality scores and the subjective scores for all training samples. The Jacobian matrix based Levenberg-Marquardt method is used for faster convergence than traditional gradient descent~\cite{stats_book}. Once the optimal $\mathcal{M}$ and $\mathcal{N}$ and corresponding feature set is determined, we use all training data (containing 52 samples) to retrain the network weights. The resulting set of weights is listed in Table~\ref{tab:weights}, and is used to predict the quality of the samples in the testing set. All input features are normalized to zero mean and unit variance.
\begin{table}[!t]
\centering
\small{
\caption{Learned Weights for Neural Network, Weights on hidden nodes are ordered based on the features listed in Table~\ref{tab:feats}, the last number for each node is the bias term for that node}
\label{tab:weights}
\begin{tabular}{|c||l|}
\hline
Node 1 & $-0.5236$, $2.8352$, $-0.6619$, \\
 & $2.2123$, $-0.2637$, $-0.3205$, \\
 & and $3.1631$ \\ \hline
Node 2 & $5.6230$, $-4.7354$, $2.1113$, \\
 & $-2.6986$, $-1.9342$, $6.0606$, \\ 
 & and $1.8050$ \\ \hline
Node 3 & $1.6702$, $-0.8454$, $1.8230$, \\
 & $-2.4986$, $1.5318$, $-0.2756$, \\
 & and $-2.2932$ \\ \hline
Output layer & $-1.2341$, $-0.5106$, $-1.1324$, \\
 & and $0.0932$ \\ \hline
\end{tabular}}
\vspace{-.1in}
\end{table}
\section{Result}\label{sec:res}
\begin{figure}[!t]
\centering
\includegraphics[width=0.75\columnwidth]{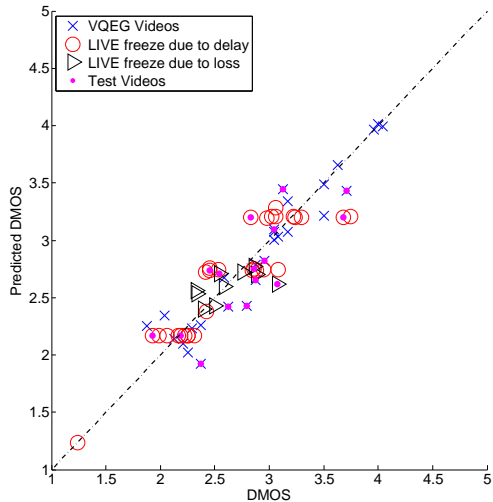}
\caption{Scatter plot for the proposed metric using the optimally selected features and the trained neural network structure. Points with ``$\cdot$'' are test videos, while others are training videos.}
\label{fig:results}
\end{figure}

\begin{figure}[!t]
\centering
\includegraphics[width=0.8\columnwidth]{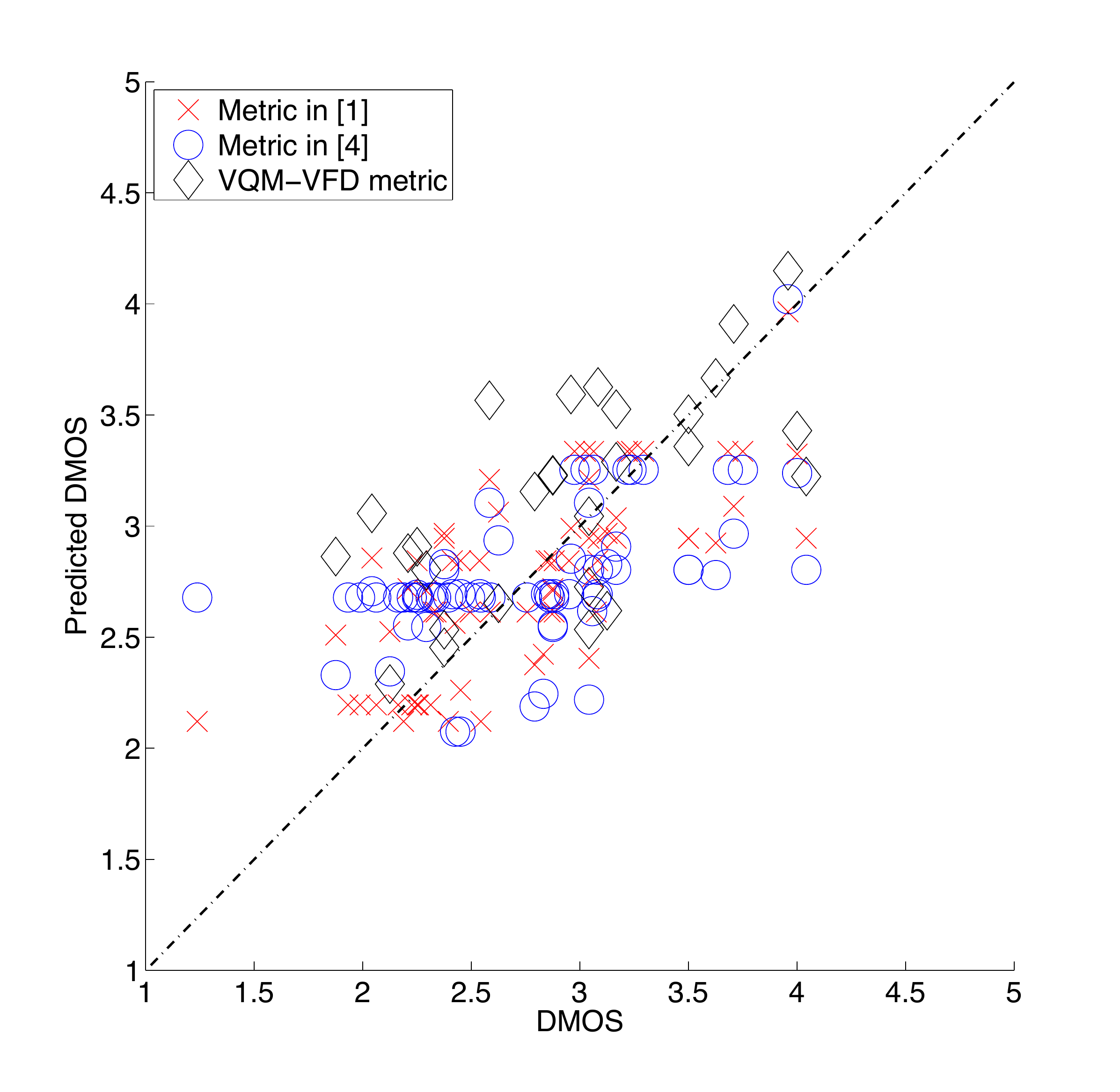}
\cprotect\caption{Scatter plot for NR Metric in~\cite{Quan_ICIP09_NRfreeze}, NR metric in~\cite{TemporalMetric_2013}, and Full-referenced \verb+VQM-VFD+ metric in~\cite{NTIA_VQM}. For~\cite{Quan_ICIP09_NRfreeze} and~\cite{TemporalMetric_2013} , results for both VQEG and LIVE datasets are reported. For~\cite{NTIA_VQM}, only results are VQEG video are included.}
\label{fig:vqmvfd}
\end{figure}

Figure~\ref{fig:results} shows the scatter plots of subjective ratings (in terms of DMOS) and the predicted scores for the proposed neural network using the optimized network structure and features. Note that the proposed metric achieves a high correlation of over 0.9 for the whole set and 0.8 for a separate testing set. 
\begin{table}[!t]
\centering
\small{
\caption{Performance comparison}
\label{tab:perfcomp}
\begin{tabular}{|c||c|c|c|}
\hline
Metrics & $\mathrm{PCC}$ & $\mathrm{SROCC}$ & $\mathrm{rRMSE}$ \\ \hline \hline
\cite{Quan_ICIP09_NRfreeze} & 0.71 & 0.73 & 8.87\% \\ \hline
\cite{TemporalMetric_2013} & 0.54 & 0.51 & 12.1\% \\ \hline
\verb+VQM-VFD+~\cite{NTIA_VQM} on VQEG videos & 0.65 & 0.62 & 10.1\% \\ \hline
Proposed (whole set) & 0.93 & 0.91 & 5.04\% \\ \hline
Proposed (training set) & 0.95 & 0.94 & 3.34\% \\ \hline
Proposed (testing set) & 0.80 & 0.79 & 6.7\% \\ \hline
\end{tabular}}
\vspace{-.1in}
\end{table}
We also show results by the NR temporal jerkiness metric in~\cite{Quan_ICIP09_NRfreeze} and~\cite{TemporalMetric_2013}, and the full reference \verb+VQM-VFD+~\cite{NTIA_VQM} metric in Fig.~\ref{fig:vqmvfd}. 

We report the Pearson correlation coefficient ($\mathrm{PCC}$), Spearman rank order correlation coefficient ($\mathrm{SROCC}$) and relative root-mean-square error ($\mathrm{rRMSE}$) of these methods in Table~\ref{tab:perfcomp}. For~\cite{Quan_ICIP09_NRfreeze}, we find the optimal parameters using the same training set used in our proposed metric. We have found that the number of bins specified for the histogram is a crucial factor for the metric performance. Because it does not include any video content characteristics, it could not differentiate videos with the same freezing pattern (as is the case for LIVE mobile database). For~\cite{TemporalMetric_2013}, we used the same definitions and corresponding weights reported for freezes happened in different parts of the video as reported in~\cite{TemporalMetric_2013}. For~\cite{NTIA_VQM}, we use the recommended full reference calibration method~\cite{NTIA_FRcal} for temporal alignment. We only report results for the VQEG videos since we are not able to run the provided software~\cite{VQMVFD} on LIVE videos. It is clear that, from both the scatter plots and Table~\ref{tab:perfcomp}, the proposed method correlates more closely to the subjective ratings than the comparison methods. Note that, we are aware that the metric proposed in~\cite{TemporalMetric_2013} is designed for longer video sequences, we observe that the limitation of their model is the main reason it performs poorly in comparison. In fact, their metric can be expressed as a one layer neural network.
\section{Conclusion}\label{sec:conc}
In this paper, we propose a novel NR temporal jerkiness quality metric for videos acting on a carefully chosen set of features. The network structure and the features are optimized through a cross validation procedure using training data extracted from publicly available annotated video datasets. 

The resulting feature set and the trained network are further evaluated on separate testing data. The proposed metric works equally well for frame freezes due to either packet loss or packet delay. Our metric extracts 6 features from the distorted video which describe the freeze duration, inter-freeze distance, and the ratio between the frame difference after a freeze and the background frame difference. The last feature depends on the motion characteristics of the video. Our data driven approach then utilizes a 1-hidden layer neural network trained by a subset coming from the publicly available annotated video datasets to find the mapping from the features to the quality. Our trained network with optimally chosen features can achieve very good performance with average $\mathrm{PCC}$ over 0.9 for the training set and achieve $\mathrm{PCC}=0.8$ for video contents not seen before. Our proposed metric significantly outperforms two previously reported no-reference video quality metrics and one full reference metric.

%





\ifCLASSOPTIONcaptionsoff
  \newpage
\fi



\bibliographystyle{IEEEtran}
{\bibliography{NR_freeze_metric}}



\begin{IEEEbiography}[{\includegraphics[width=1in,height
=1.25in,clip,keepaspectratio]{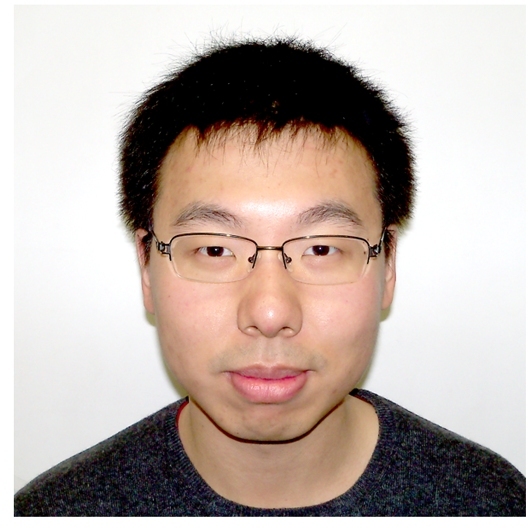}}]{Yuanyi Xue}
 (S’09) received the B.Eng.(’08) degree in Electrical Engineering from Southeast University, Nanjing, China. He received the M.S. degree in Electrical Engineering from Polytechnic School of Engineering of New York University (then Polytechnic Institute of NYU) in 2010. He is now a Ph.D. candidate major in Electrical Engineering in Polytechnic School of Engineering of New York University. His research interests include perceptual video quality assessment and modeling, sparse signal recovery, and machine learning applications for video signal processing.
\end{IEEEbiography}
\begin{IEEEbiography}[{\includegraphics[width=1in,height
=1.25in,clip,keepaspectratio]{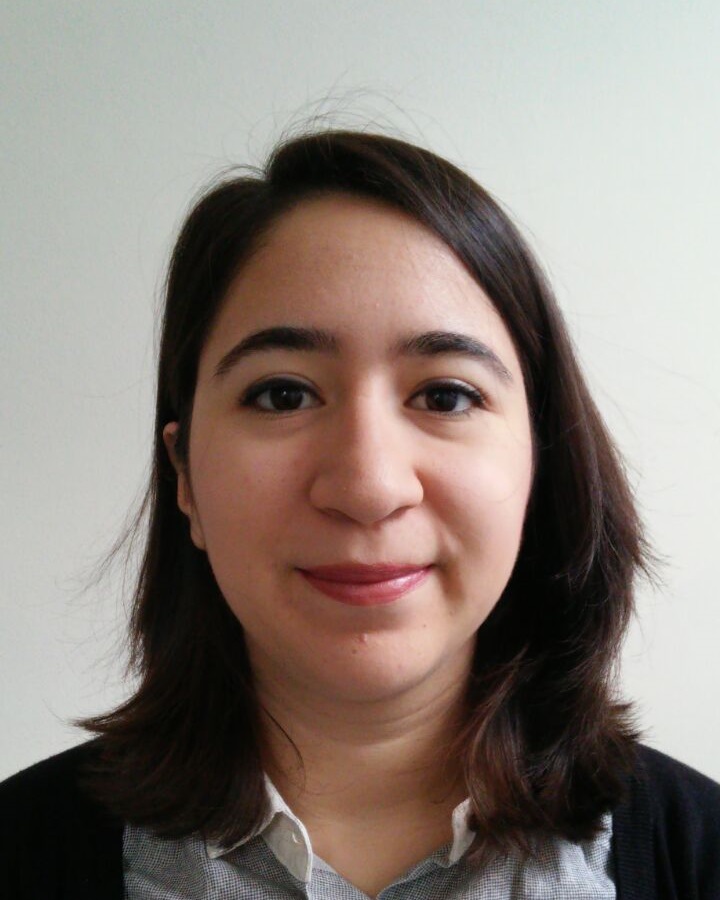}}]{Beril Erkin}
 is a Video Software engineer at Dialogic Inc, New Jersey. She received her Bachelor’s degree with honors in Electrical and Electronics Engineering at Middle East Technical University in Ankara, Turkey, before completing a Master’s degree in Electrical Engineering at Polytechnic School of Engineering of New York University in 2014. Her research interests include video communications, WebRTC, image analysis, computer vision, and machine learning. 
\end{IEEEbiography}
\begin{IEEEbiography}[{\includegraphics[width=1in,height
=1.25in,clip,keepaspectratio]{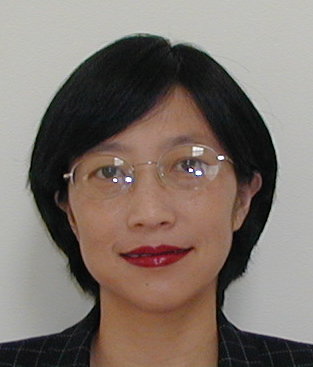}}]{Yao Wang}
 (M’90-SM’98-F04) received the B.S. and M.S. degrees in Electronic Engineering from Tsinghua University, Beijing, China, in 1983 and 1985, respectively, and the Ph.D. degree in Electrical and Computer Engineering from University of California at Santa Barbara in 1990. Since 1990, she has been on the faculty of Electrical and Computer Engineering, Polytechnic School of Engineering of New York University (formerly Polytechnic University, Brooklyn, NY). She was on sabbatical leave at Princeton University in 1998 and at Thomson Corporate Research, Princeton, in 2004-2005. She was a consultant with AT\& T Labs - Research, formerly AT\& T Bell Laboratories, from 1992 to 2000. Her research areas include video communications, multimedia signal processing, and medical imaging. She is the leading author of a textbook titled Video Processing and Communications, and has published over 150 papers in journals and conference proceedings. She has served as an Associate Editor for \textsc{IEEE Transactions on Multimedia} and \textsc{IEEE Transactions on Circuits and Systems for Video Technology}. She received New York City Mayor’s Award for Excellence in Science and Technology in the Young Investigator Category in year 2000. She was elected Fellow of the IEEE in 2004 for contributions to video processing and communications. She is a co-winner of the IEEE Communications Society Leonard G. Abraham Prize Paper Award in the Field of Communications Systems in 2004, and a co-winner of the IEEE Communications Society Multimedia Communication Technical Committee Best Paper Award in 2011. She was a keynote speaker at the 2010 International Packet Video Workshop. She received the Overseas Outstanding Young Investigator Award from the Natural Science Foundation of China in 2005 and was named Yangtze River Lecture Scholar in Tsinghua University by the Ministry of Education of China in 2007.
\end{IEEEbiography}
\end{document}